\documentclass[article, nojss]{jss}

\usepackage{thumbpdf,lmodern,amsmath}
\usepackage[T1]{fontenc}
\usepackage{graphics, graphicx}

\usepackage{algpseudocode}
\usepackage[plain]{algorithm}
\usepackage{algorithmicx}

\makeatletter
\newcommand\fs@truled{%
  \def\@fs@cfont{\normalfont}%
  \let\@fs@capt\floatc@plain
  \def\@fs@pre{\hrule\kern2pt}%
  \def\@fs@post{}%
  \def\@fs@mid{\hrule\kern2pt}%
  \let\@fs@iftopcapt\iftrue
}
\makeatother
\floatstyle{truled}
\restylefloat{algorithm}

\author{
   Adrian E. Raftery\\University of Washington
   \And Hana \v{S}ev\v{c}{\'{\i}}kov{\'a}\\University of Washington
   \And Bernard W. Silverman\\University of Nottingham}
\Plainauthor{Adrian E. Raftery,  Hana \v{S}ev\v{c}{\'{\i}}kov{\'a}, 
Bernard W. Silverman}

\title{The \pkg{vote} Package: Single Transferable Vote and Other Electoral Systems in \proglang{R}}
\Plaintitle{Single Transferable Vote and Other Electoral Systems in R}
\Shorttitle{STV and Other Electoral Systems in \proglang{R}}

\Abstract{
We describe the \pkg{vote} package in \proglang{R}, which implements
the plurality (or first-past-the-post), two-round runoff, score, approval and single transferable
vote (STV) electoral systems, as well as methods for selecting the Condorcet
winner and loser. We emphasize the STV system,
which we have found to work well in practice for multi-winner elections
with small electorates, such as committee and council elections, and the
selection of multiple job candidates. For single-winner elections, the STV 
is also called instant runoff voting (IRV), ranked choice voting (RCV), 
or the alternative vote (AV) system.  The package also implements the
STV system with equal preferences, for the first time in a software package,
to our knowledge. It also implements a new variant of STV, in which a minimum number of candidates from a specified group are required to be elected.  We illustrate the package with several real examples.
}

\Plainkeywords{}

\Address{
}

\begin{document}



\section{Introduction} 
\label{sec:intro}
The \pkg{vote} package implements several electoral methods: plurality
voting, approval voting, score voting, Condorcet methods, and 
single transferable vote (STV) methods \citep{vote}.

In developing the package, we were motivated particularly by the needs of
organizations with small electorates, such as learned societies, clubs
and university departments, who may need to elect more than one person
in a given election. In the early 1980s, one of us (BWS) was a member of the 
Royal Statistical Society (RSS) Council. At that time, 
six members of the Council were elected at a time. A nominating committee
nominated six candidates and the RSS membership as a whole voted, with each
member allowed to vote for up to six candidates, and the six candidates with 
the most votes being elected. Usually there were only 
the six nominated candidates, but that year a seventh candidate 
stood on a platform different from that of the ``official'' candidates.
This candidate received votes from about a quarter of the electorate, but was not 
elected because the other three-quarters of the members voted as a block for
the six candidates proposed by the nominating committee.

This was viewed as unsatisfactory, because the seventh candidate's position
was not represented on the Council, even though it had substantial support
among the RSS membership.
This led the RSS Council to undertake a study of electoral methods for
multi-winner elections, with a view to adopting a more representative system.
They selected the Single Transferable Vote (STV) method, which was then 
adopted for Council elections, initially using a program in the 
\proglang{Pascal} programming language developed by \citet{Hill&1987}.
In the next election, held under STV, the seventh candidate stood again, 
and was elected. STV has been used since then to elect the RSS Council.

In 2002, the Institute of Mathematical Statistics (IMS), the leading 
international association of academic mathematical statisticians, 
considered the same issue and came to the same conclusion, also adopting STV
for its Council elections. They used an \proglang{R}  program developed by BWS
\citep{Silverman2002,Silverman2003}, who was also then the IMS President.
This \proglang{R} program became the core of the \pkg{vote} package that
we are describing here.  This STV electoral method has been used since 
then by the IMS. 

Since then, another one of us (AER) has implemented the STV method 
in the context of small electorates selecting or ranking multiple 
candidates, such as nominating committees selecting multiple awardees 
for a prize, or academic departments selecting job candidates for interviews.
Those involved have generally reported finding the method satisfactory.
This experience has led to several modifications of the program that 
are also implemented in the package.

Our implementation and discussion of STV and other systems is aimed particularly at those involved in non-party-political elections and decisions, such as those outlined above. Questions of what approaches are or are not desirable for national elections are matters of political science beyond the scope of this paper, which is not intended to advocate for or against the use of any particular voting systems in that context; however a brief review may be informative.

The USA and the UK, for their national legislatures, almost entirely use the plurality,  ``first past the post'' or ``winner takes all'' system, where the leading candidate in each district is elected.   The Electoral College for the US presidency is also elected this way, but with an election between slates rather than individuals, in all states except Maine and Nebraska.  On the other hand, the majority of countries use some system that (in principle at least) aims for the elected body to represent proportionately the views of the wider electorate, either over the country as a whole or within larger electoral districts. However, pure proportional systems are fairly unusual, for example because in nationwide proportional systems there is often a threshold below which a party will not have any representation. The Single Transferable Vote system is used to elect the parliaments or national assemblies of the Republic of Ireland, Northern Ireland and Malta, as well as upper houses and/or local assemblies in some other countries (Wikipedia 2020c), and we draw an example from a Dublin election in the paper. 

As we have said, it is not our purpose to advocate any one electoral method, and indeed it
is well known that there is no one method that dominates all others given
a reasonable set of criteria, according to the impossibility theorems of
\citet{Arrow1963}, \citet{Gibbard1973} and \citet{Satterthwaite1975}.
Indeed, method choice can depend on the purpose of the election, and 
a method that works well for one purpose (such as representing the views
of the electorate), may not be best for others (such as electing an effective
team to work together) \citep{Syddique1988}.
As a result, we have implemented multiple electoral methods in the package. 
Pros and cons of a wide range of different electoral systems 
are described in \citet{Ace2020}, but these focus on nationwide political
elections, whereas here we also pay attention to smaller, often non-political
elections, such as those for councils and committees.

The paper is organized as follows. In Section \ref{sec:methods} we describe
the plurality, two-round runoff,  approval, score and Condorcet vote-counting methods.
In Section \ref{sec:STV} we describe the STV method, including the first 
software implementation of the equal preference STV method, to our knowledge.
This also describes a new variant of STV which enforces minimal representation
of a marked group.
In Section \ref{sec:examples} we describe three multi-winner elections 
with electorates of different sizes: 
an election from one constituency in the 2002 Irish General Election,  
an election of the IMS Council, and a vote to select job candidates 
by a university department.
We conclude in Section \ref{sec:discussion} with discussion of issues 
including other \proglang{R} packages for vote-counting.

\section{Electoral Methods}
\label{sec:methods}
In this section, we describe several electoral methods and how they are 
implemented in the \pkg{vote} package. We defer description of STV to
Section \ref{sec:STV}.

We first illustrate the results here with the toy \code{food_election} dataset:
\begin{CodeInput}
R> library (vote)
R> data (food_election)
R> food_election
\end{CodeInput}
\begin{CodeOutput}
   Oranges Pears Chocolate Strawberries Sweets
1       NA    NA         1            2     NA
2       NA    NA         1            2     NA
3       NA    NA         1            2     NA
4        2     1        NA           NA     NA
5       NA    NA        NA            1     NA
6        1    NA        NA           NA     NA
7       NA    NA        NA           NA      1
8        1    NA        NA           NA     NA
9       NA    NA         1            2     NA
10      NA    NA         1           NA      2
11       1    NA        NA           NA     NA
12      NA    NA         1            2     NA
13      NA    NA         1            2     NA
14      NA    NA         1            2     NA
15      NA    NA         1           NA      2
16       1    NA        NA           NA     NA
17      NA    NA         1           NA      2
18       2     1        NA           NA     NA
19      NA    NA         1           NA      2
20      NA    NA         1            2     NA
\end{CodeOutput}
In this toy dataset, voters were asked to rank the options in order of preference.
They gave only their first two preferences although they could have given more; an \code{NA} indicates that no preference was expressed.

\subsection{Plurality Voting}
\label{subsec:plurality}
Plurality voting, or First-Past-The-Post,
 is used for single-winner elections, such as 
elections to the House of Representatives in the USA or the House of 
Commons in the UK. Each voter votes for one candidate and the candidate
with the most votes wins. 

To implement this with our toy dataset, we first converted it to a dataset 
where only first preferences count:
\begin{CodeInput}
R> food_election_plurality <- 1 * (food_election == 1 & !is.na (food_election))
R> head(food_election_plurality)
\end{CodeInput}
\begin{CodeOutput}
     Oranges Pears Chocolate Strawberries Sweets
[1,]       0     0         1            0      0
[2,]       0     0         1            0      0
[3,]       0     0         1            0      0
[4,]       0     1         0            0      0
[5,]       0     0         0            1      0
[6,]       1     0         0            0      0
\end{CodeOutput}

We then counted the votes using the \code{plurality} command:
\begin{CodeInput}
R> plurality (food_election_plurality)
\end{CodeInput}
\begin{CodeOutput}
Results of Plurality voting
===========================                           
Number of valid votes:   20
Number of invalid votes:  0
Number of candidates:     5
Number of seats:          1


|    |Candidate    | Total| Elected |
|:---|:------------|-----:|:-------:|
|1   |Chocolate    |    12|    x    |
|2   |Oranges      |     4|         |
|3   |Pears        |     2|         |
|4   |Strawberries |     1|         |
|5   |Sweets       |     1|         |
|Sum |             |    20|         |

Elected: Chocolate 
\end{CodeOutput}

Plurality voting has the advantage of simplicity. 
In political elections, it tends not to yield results which are in direct proportion to support among the voters, but to amplify pluralities when compared to proportional voting systems which merge single-winner districts into larger multi-member groups.  In general, any large party which has strong support across a large number of electoral districts will do well under plurality voting, while smaller parties or interests will tend to be underrepresented numerically, especially if they are evenly or thinly spread.  This may mean that important interests are not represented, while on the other hand it may present a barrier to the traction of extremist groups.  The US Electoral College is, in nearly all states, elected by a plurality voting system, with multiple members all being elected simultaneously.  

Plurality voting in individual-member districts tends to lead to one-party governments with working majorities, even when the leading party does not achieve half of the popular vote. It also allows districts to be smaller to facilitate direct contact between a voter and their representative, and identifies each representative more closely with all the voters in their district.  

Another effect of plurality voting can be to ``waste'' the votes of those who live in highly polarised districts, because they win their particular district by a very wide margin; this seems to be a deliberate feature of much redistricting in the USA.  
In non-political elections in the smaller contexts of primary concern in this paper, there is little or no need for a stable one-``party'' result, and the desirability of closer proportional representation of the views of the voters is less contentious, and so there is likely to be a clearer case for using other voting systems wherever possible.   

\subsection{Two-round Runoff Voting}
\label{subsec:two-round}
Two round systems are also used for single-winner elections.
In the first round, voters vote for their first preference.
If no candidate gets a majority, there is a second round that involves the
top two candidates. Voters vote again, and the candidate getting more 
votes wins.

In the \pkg{vote} package, we implemented a variant of this system that
can be counted in a single pass over the data.
Each voter ranks the candidates in order of preference. 
The first round takes place as described. The second round is counted
as if voters voted for the remaining candidate for which they had a higher preference.

To illustrate the two-round runoff system, we modify the food election data
by removing voters 12--15, so that Chocolate does not have a majority on the 
first round:

\begin{CodeInput}
R> food_election3 <- food_election[-c(12:15),]
R> tworound.runoff (food_election3)
\end{CodeInput}

\begin{CodeOutput}
Results of two-round-runoff voting
==================================                           
Number of valid votes:   16
Number of invalid votes:  0
Number of candidates:     5
Number of seats:          1


|    |Candidate    | Total| Percent| ROTotal| ROPercent| Elected |
|:---|:------------|-----:|-------:|-------:|---------:|:-------:|
|1   |Oranges      |     4|    25.0|       6|      42.9|         |
|2   |Pears        |     2|    12.5|       0|       0.0|         |
|3   |Chocolate    |     8|    50.0|       8|      57.1|    x    |
|4   |Strawberries |     1|     6.2|       0|       0.0|         |
|5   |Sweets       |     1|     6.2|       0|       0.0|         |
|Sum |             |    16|   100.0|      14|     100.0|         |

Elected: Chocolate 
\end{CodeOutput}

We see that no candidate got a majority on the first round, although
Chocolate came close. In the second round, the two top vote-getters,
Chocolate and Oranges, squared off, and Chocolate won. 

In the \code{tworound.runoff} function, a tie in either the first 
or the runoff round is resolved by random draw. 
A random seed can be specified, so that the results are replicable. 

Two-round elections are quite common, most famously for French presidential
elections since 1965. In practice it is usually carried out by voters
actually voting twice, rather than ranking candidates as here. An exception to it is a special case of  the two-round runoff, called supplementary voting where voters give only their first and second preferences on one ballot, the same way as happened in our food example. Supplementary voting is used for example in electing mayors in England including the Mayor of London~\citep{LondonElects2020}. 

The two-round runoff system differs from plurality voting in that voters for candidates
with low levels of support can change their votes to one of the  leading candidates, so that they can express support for a possibly less popular first choice without their vote being ``wasted''.  Of course, the choice between the two finalists shares some of the aspects of plurality voting.

\subsection{Approval Voting}
\label{subsec:approval}
Approval voting was advocated by \citet{BramsFishburn1978}. 
In this system, voters vote for as many candidates as they wish. 
It has been most often advocated for single-winner elections, in which case
the winner is the candidate with the most votes \citep{BramsFishburn2007}.
A direct extension to multi-winner elections with $m$ winners is that voters 
vote in the same way, and the $m$ candidates with the most votes win. 

Counting the votes is simple. The argument \code{mcan} determines the number of winners $m$:
\begin{CodeInput}
R> food_election_approval <- 1 * !is.na (food_election)
R> approval (food_election_approval, mcan = 2)
\end{CodeInput}
\begin{CodeOutput}
Results of Approval voting
==========================                           
Number of valid votes:   20
Number of invalid votes:  0
Number of candidates:     5
Number of seats:          2


|    |Candidate    | Total| Elected |
|:---|:------------|-----:|:-------:|
|1   |Chocolate    |    12|    x    |
|2   |Strawberries |     9|    x    |
|3   |Oranges      |     6|         |
|4   |Sweets       |     5|         |
|5   |Pears        |     2|         |
|Sum |             |    34|         |

Elected: Chocolate, Strawberries 
\end{CodeOutput}

Approval voting for multi-winner elections has been criticized on various
grounds, e.g. \citet{Hill1988}, and indeed in the book by 
\citet{BramsFishburn1983} that advocated and popularized
approval voting for single-winner elections. 
For elections in which there
are parties or slates of candidates, it would tend to lead to the 
election of all the members of the most supported party or slate,
as happened in the RSS Council election that first motivated this work.
However, one of us [AER] has 
participated in multi-winner elections using approval voting and has observed
it to work well, particularly when there are many candidates about whom
information is limited, and there are no parties or slates. 
One example could be the early stages of job candidate selection, 
when a long list is being whittled down to a small set of finalists.

\subsection{Score Voting}
\label{subsec:score}
In score, or range voting, each voter gives each candidate a score within
a prespecified range. If the voter does not give a score to a particular
candidate, a corresponding prespecified score is assigned. The candidates
with the lowest scores win (or the highest scores if higher scores are better). 
In the \code{score} function, the argument \code{larger.wins} specifies
whether lower scores are better or higher scores are better. The argument \code{max.score} sets the prespecified non-vote score.
Here we illustrate score voting by applying it to the food election example, where 
the score is equal to the preference, a non-vote is assigned a value of 6,
and lower scores are better:

\begin{CodeInput}
R> score (food_election, larger.wins = FALSE, mcan = 2, max.score = 6)
\end{CodeInput}

\begin{CodeOutput}
Results of Score voting
=======================                           
Number of valid votes:   20
Number of invalid votes:  0
Number of candidates:     5
Number of seats:          2


|    |Candidate    | Total| Elected |
|:---|:------------|-----:|:-------:|
|1   |Chocolate    |    60|    x    |
|2   |Strawberries |    83|    x    |
|3   |Oranges      |    92|         |
|4   |Sweets       |    99|         |
|5   |Pears        |   110|         |
|Sum |             |   444|         |

Elected: Chocolate, Strawberries 
\end{CodeOutput}

Score voting is often used by committees for purposes such as selecting
grant applications to be funded.
In such cases there are often many candidates or applications to be assessed, 
and it would not be feasible for the voters to produce a complete ranking. 
Score voting is similar to multi-winner approval voting
in this sense, but allows for more refined assessment by the voters.
Multi-winner approval voting is actually a special case of score voting.

\subsection{Condorcet Method}
\label{subsec:Condorcet}
The Condorcet method is attributed to Marquis de Condorcet~\citep{Condorcet1785}. It is a single-winner method where voters rank the candidates according to their preferences. The so-called {\em Condorcet winner} is the candidate who wins the majority of votes in all head-to-head comparisons. In other words, each candidate is compared pairwise to all other candidates. To become the Condorcet winner one has to win all such comparisons. Analogously, a {\em Condorcet loser} is the candidate who loses in every pairwise comparison.

The \code{condorcet} function can be applied directly to the food election data:
\begin{CodeInput}
R> condorcet(food_election)
\end{CodeInput}

{\small
\begin{CodeOutput}
Results of Condorcet voting
===========================                           
Number of valid votes:   20
Number of invalid votes:  0
Number of candidates:     5
Number of seats:          1


|             | Oranges| Pears| Chocolate| Strawberries| Sweets| Total| Winner | Loser |
|:------------|-------:|-----:|---------:|------------:|------:|-----:|:------:|:-----:|
|Oranges      |       0|     1|         0|            0|      1|     2|        |       |
|Pears        |       0|     0|         0|            0|      0|     0|        |   x   |
|Chocolate    |       1|     1|         0|            1|      1|     4|   x    |       |
|Strawberries |       1|     1|         0|            0|      1|     3|        |       |
|Sweets       |       0|     1|         0|            0|      0|     1|        |       |

Condorcet winner: Chocolate
Condorcet loser: Pears
\end{CodeOutput}
}

The output above shows the results of all the pairwise comparisons. Chocolate beat all other candidates and was therefore the Condorcet winner. Similarly, Pears lost against all other candidates and was thus the Condorcet loser. 

The Condorcet method does not guarantee that a Condorcet winner exists. There are many different ways to deal with such a situation, see for example~\citet{WikipediaCondorcet2020}. Our implementation offers the possibility of a run-off (argument \code{runoff}). In this case, two or more candidates with the most pairwise wins are selected and the Condorcet method is applied to such subset. If more than two candidates are in such run-off, the selection is performed repeatedly, until either a winner is selected or no more selection is possible. 

To our knowledge, the Condorcet method is not used for governmental elections anywhere in the world. \citet{WikipediaCondorcet2020} cites a few private organizations that use the method, e.g., the Student Society of the University of British Columbia.

\section{Single Transferable Vote (STV)}
\label{sec:STV}
The Single Transferable Vote (STV) system is also referred to as 
Ranked Choice Voting (RCV), Instant Runoff Voting (IRV), or the Alternative
Vote (AV) system for single-winner elections, 
and as Multi-Winner Ranked Choice Voting for multi-winner elections.
One of the properties of the Single Transferable Vote system is that if any subset of candidates gets a sufficient share of the votes, anything strictly exceeding $1/(m+1)$, where $m$ is the number of candidates to be elected,  then one of this group is bound to be elected. 
To be precise, what is required is that a proportion above $1/(m+1)$ of the voters have to put all the candidates in the subset at the top of their list of preferences, but it does not matter in what order.
This would apply equally if the subset was a particular slate/party, or specified by some other group characteristic such as sex or race or geographical location or career stage or subject area, even if the subset was not consciously constituted.  
In particular, if a candidate's proportion of the first preference votes is above $1/(m+1)$, then that candidate will be successful.

There is also the fact that a group is not disadvantaged if more of its members stand for election.  Unlike in some other systems, they cannot cancel each other out. 

When STV was adopted for the elections of the Council of the RSS in the mid-1980s and the IMS in 2002, it was hoped that it would lead to more diverse Councils than the results of the previous methods, and also that individual members, other than those chosen by the nominating committee, would feel able to stand with a real chance of being elected.

\subsection{STV Method}
\label{subsec:STV}
There are many descriptions of the STV system \citep{Newland&1997,FairVote2020}, and its history \citep{Hill1988,Tideman1995}.
The basic principle is that voters rank the candidates in order of preference. 
In order to be elected a candidate must achieve the quota of $N/(m+1) + \varepsilon$, where $N$ is the total number of votes cast, $m$ is the number of candidates to be elected (or seats), and $\varepsilon$ is a pre-specified small positive number, often taken to be 1 when the electorate is large and 0.001 when it is small. Excess votes over the quota are appropriately downweighted and allocated to the next preference of voters. If no candidate reaches the quota, the candidate with the smallest number of votes is eliminated and his or her votes are transferred to the next preferences.

Voters are asked to rank the candidates $1,2,3,\ldots$ until they have no further preference between candidates. Thus 1 is a voter's first preference, 2 is their next choice, and so on. There is no disadvantage to higher candidates in expressing a full list of preferences; later preferences are used only when the fate of candidates given higher preferences has been decided one way or the other. 

By default, a vote is considered spoiled if the preferences are not numbered consecutively starting at 1. However, if this is not desired, the votes can be preprocessed to be consecutive using the \code{correct.ranking} function in the \pkg{vote} package. A useful application of this correction is the case when a candidate has to be removed, perhaps because of having withdrawn his or her candidacy. In this case, the function \code{remove.candidate} can be used, which removes the given candidate(s) from the set of votes, and also adjusts the preferences to be consecutive. 

Also by default, apart from the candidates not numbered at all, no ties are allowed among the numbered preferences. 
However, equal preferences can be allowed by using the setting
\code{equal.ranking=TRUE} in the \code{stv} function,
as described in more detail in Section \ref{subsec:equalpref}, 

The fact that some voters may not express a full list of preferences can be
allowed for by reducing the quota in later counts\footnote{In STV,
the process of distributing the surplus or votes of a candidate who is elected
or eliminated is referred to variously as a count, a stage or a round. 
Here we use the term count. The tabulation of the first preference votes
is then called the first count.}.
In the \pkg{vote} package, the default is that the quota is reduced in
later counts. However in some STV systems (such as the electoral system in
the Republic of Ireland), the quota remains constant over counts at the value
that is initially defined. This is specified in the \pkg{vote} package by
setting the argument ``\code{constant.quota = TRUE}'' in the \code{stv} 
function. In this implementation of STV, the last candidate is often 
elected without reaching the quota, which does not happen when the quota
is reduced appropriately at each count.

In the \pkg{vote} package, the votes should be entered into a matrix or data frame, with the
header containing the names of the candidates and each row the votes cast, with blank
preferences being replaced by zeroes or \code{NA}s. This will often be done by entering the votes into
a spreadsheet first and then reading the spreadsheet into \proglang{R}. If the data are stored in a text file, the package allows one to pass the name of the file directly into the \code{stv} function while setting the column separator in the \code{fsep} argument. 

At the end of the process, the program yields a list of the successful candidates in the order in which they were elected. It also usually yields a complete ordering
of the candidates. This may be useful, for example, if the purpose of the
election is to select job candidates, and one wishes to have an ordered list
of the initially unsuccessful candidates in case any of those selected
decline the offer.

Until the 1980s, STV elections were counted manually by physically transferring
ballot papers from the pile of the candidate being elected or
eliminated, to those of the benefitting candidates.
This remains the case in several long-established STV election systems.
\citet{Meek1969,Meek1970} described the form a computer-based STV
system could take, and this was implemented in \proglang{Pascal} by \citet{Hill&1987}. 
This code was used for the RSS Council elections.
A modified version was implemented in \proglang{R} by 
\citet{Silverman2002,Silverman2003}, and this was the 
starting point for the current STV implementation in the \pkg{vote} package.

Here is the result of the food election with two candidates to be elected,
 using the \code{stv} function:
\begin{CodeInput}
R> stv (food_election, mcan = 2)
\end{CodeInput}
\begin{CodeOutput}
Results of Single transferable vote
===================================                           
Number of valid votes:   20
Number of invalid votes:  0
Number of candidates:     5
Number of seats:          2


|             |         1| 2-trans|     2| 3-trans|      3| 4-trans|       4|
|:------------|---------:|-------:|-----:|-------:|------:|-------:|-------:|
|Quota        |     6.668|        | 6.667|        |  6.667|        |   5.278|
|Oranges      |     4.000|   0.000| 4.000|       2|  6.000|   0.000|   6.000|
|Pears        |     2.000|   0.000| 2.000|      -2|       |        |        |
|Chocolate    |    12.000|  -5.332|      |        |       |        |        |
|Strawberries |     1.000|   3.555| 4.555|       0|  4.555|   0.000|   4.555|
|Sweets       |     1.000|   1.777| 2.777|       0|  2.777|  -2.777|        |
|Elected      | Chocolate|        |      |        |       |        | Oranges|
|Eliminated   |          |        | Pears|        | Sweets|        |        |

Elected: Chocolate, Oranges 
\end{CodeOutput}

Oranges was elected second, whereas under the approval vote system with
first and second preferences treated equally, Strawberries was elected second.
This reflects the fact that Oranges had 4 first preferences whereas 
Strawberries had only 1. Under STV, a vote is credited entirely to the
first preference candidate unless that candidate is elected or eliminated,
in which case the second preferences come into play. 
Strawberries had 8 second-preference votes, all of which were from voters
who voted for Chocolate first. The quota was only 56\% of the votes for
Chocolate, and so 44\% of Chocolate's votes were transferred when 
Chocolate was elected. Strawberries gained 3.555 votes this way from
its second preference votes, but this was not quite enough to overcome
Orange's advantage in first preferences.
The complete ordering of candidates can be read off the results: 
Chocolate, Oranges, Strawberries, Sweets, Pears. Setting the argument \code{complete.ranking} to \code{TRUE} will include the complete ordering as part of the output.

The package has several functions for visualizing the STV results, and we will illustrate these in the Examples section below. In addition, \code{summary} functions are available for the resulting objects of all voting methods in the package. In the case of \code{stv}, the \code{summary} function returns a data frame containing the table shown in the above output which can be used for further processing, for example for storing in a spreadsheet.

\subsection{Computational Methods}
\label{subsec:computational}
The algorithm used for counting STV elections using the \code{stv}
function in the \pkg{vote} package is shown in Algorithm \ref{alg:stv}. 
There are only two changes needed to implement STV with equal preferences;
these are shown in Section \ref{subsec:equalpref}.

\begin{algorithm}[!htp]
    \caption{STV algorithm. The input data consist of a matrix $X$ of the votes of size $N\times M$, with $N$ being the number of ballots and $M$ the number of candidates. $m$ is the number of seats to be filled, and $\varepsilon$ is a small number used for defining the quota. 
}
    \label{alg:stv}
    \begin{algorithmic}[1] 
        \Procedure{stv}{$X, m, \varepsilon$}
        	   \State $D \gets \{1, 2, \dots, M\}$ \Comment{Set of {\it hopeful} candidates}
	    \State $E \gets \{\}$ \Comment{Set of {\it elected} candidates}
	     \State $F \gets \{\}$ \Comment{Set of {\it eliminated} candidates}
            \State $L \gets m$ \Comment{Remaining number of seats}
            \State $Y \gets X$ \Comment{Remaining votes}
            \State $c \gets 0$ \Comment{Which Count we are at}
            \State $w_i \gets 1 \quad \forall \, i = 1, \dots, N$  \Comment{Initialize a vector of weights, one per voter}
            \While{$L > 0$} \Comment{End if there are no remaining seats}
               \State $c \gets c + 1 $ \Comment{Increase Count}
                \State\label{algline:u} $u_{i,j} \gets w_i \delta_{Y_{i,j}}(1)  \quad \forall \,  i = 1, \dots, N, \; j = 1,\dots,M$ \Comment{Weighted first preferences}
                \State $v_{c,j} \gets \sum_{i=1}^N  u_{i,j} \quad \forall \, j = 1,\dots,M$ \Comment{Sum of weighted first preferences}
                \State $Q \gets \sum_{j=1}^M v_{c,j} / (L + 1) + \varepsilon $ \Comment{Compute quota}
                \If{$\max_{j\in D} v_{c,j} \geq Q$} \Comment{A candidate is to be elected}
                		\State $k \gets \arg \max_{j\in D} v_{c,j}$ \Comment{Which candidate has the most votes}
			\If{$||k|| > 1$} \Comment{If there is more than one such candidate}
			\State $k \gets $ resolve.tie.for.election($k, X, v$)  \Comment{Break tie}
			\EndIf
			\State $S \gets (\max_{j\in D} v_{c,j} - Q)/\max_{j\in D} v_{c,j}$ \Comment{Compute surplus}
			\State\label{algline:w} $w_r \gets u_{rk} * S \quad \forall \, r \text{ where } Y_{r,k} = 1$ \Comment{Recompute weights}
			\State $L \gets L - 1$ \Comment{Decrease number of available seats}
			\State $E \gets E \cup \{k\}$ \Comment{Candidate $k$ is elected}
		\Else \Comment{A candidate is to be eliminated}
			\State $k \gets \arg \min_{j\in D} v_{c,j}$ \Comment{Which candidate has the least votes}
			\If{$||k|| > 1$} \Comment{If there is more than one such candidate}
			\State $k \gets $ resolve.tie.for.elimination($k, X, v$)  \Comment{Break tie}
			\EndIf
			\State $F \gets F \cup \{k\}$ \Comment{Candidate $k$ is eliminated}
		\EndIf
		\State $D \gets D \backslash \{k\}$ \Comment{Candidate $k$ is removed from the pool of hopefuls}
		\State $Y_{i,r} \gets Y_{i,r} - 1 \quad \forall \, i = 1,\dots,N \text{ where } Y_{i,k} > 0 \text{ and } r = 1,\dots, M \text{ where } Y_{i,r} > Y_{i,k}$
		\State \Comment{Above: shift votes for voters who voted for candidate $k$}
		\State $Y_{i,k} \gets 0 \quad \forall \, i = 1,\dots,N$ \Comment{Remove votes for candidate $k$}
            \EndWhile\label{stvendwhile}
            \State {\bf return}($E, F, v$)
        \EndProcedure
    \end{algorithmic}
    \rule[0.8\baselineskip]{\linewidth}{0.4pt}\vspace{-\baselineskip}
Note: $\delta_{Y}(1) = 1$ if $Y=1$ and 0 otherwise, is the Kronecker delta function; the arg max and arg min functions return sets, with more than one element when there is a tie; and $||k||$ is the number of elements in the set $k$.
\end{algorithm}

\subsection{Tie-breaking}
\label{subsec:tie}
Suppose that on a given count, no candidate is elected and a candidate needs
to be selected for elimination, and that two or more candidates are tied
with the smallest number of votes. Then a method is needed for choosing
the one to be eliminated. The same issue arises when two candidates can be elected on the same count
with the same number of votes, namely which surplus to transfer first. 

Several different methods have been proposed. The Electoral Reform Society,
one of the leading organizations advocating the use of STV,
recommends using the Forwards Tie-Breaking Method 
\citep[Section 5.2.5]{Newland&1997}. Other methods such as Backwards Tie-Breaking,
Borda Tie-Breaking, Coombs Tie-Breaking or a combinations of those have been proposed, see e.g., 
\citet{ONeill2004, Kitchener2005, Lundell2006}.

By default the \pkg{vote} package uses the Forwards Tie-Breaking Method. This consists of eliminating/electing the
candidate who had the fewest/most votes on the first count, or on the earliest
count where they had unequal votes. If the argument \code{ties} in the \code{stv} function is set to "\code{b}", the Backwards Tie-Breaking Method is used. In this case, it eliminates/elects the candidate who has the fewest/most votes on the latest count where the tied candidates had unequal votes. 

There is no guarantee that a tie will be broken by either the Forwards or Backwards Tie-Breaking Method. Also, if one of these two methods does not break the tie, the other will not either, because the tied candidates will have the same number of votes in all the counts so far. 
In particular, this will be the case whenever a tie has to be broken on the 
first count, and it is also relatively likely when a tie arises on the 
second count.

When there is a tie that Forwards and Backwards Tie-Breaking fail to break, the \code{stv} function uses a method that compares the candidates on the basis of the numbers of individual preferences. We call this the {\em Ordered} method as it creates an ordering of the candidates before the STV count begins. First, candidates are ordered by the number of first preferences. Any ties are resolved by proceeding to the total number of second preferences, then the third preferences, and so on. If a tie cannot be resolved even by counting the last preference, then it is broken by a random draw with equal probabilities for the tied candidates. A random seed is specified so that the result is replicable.

Combining Forwards and Backwards Tie-Breaking with the Ordered method and random sampling, 
each tie in the \code{stv} function is broken in one of the following three ways:
\begin{enumerate}
\item Forwards (``f'') or Backwards (``b'')  Tie-Breaking method alone
\item Forwards or Backwards Tie-Breaking followed by the Ordered  method (``fo'', ``bo'')
\item 
Forwards or Backwards Tie-Breaking followed by the Ordered  method, and finally
random sampling (``fos'', ``bos'')
\end{enumerate}
The abbreviation of these three possibilities in parentheses is included in the STV output whenever a tie is broken during the election count.

Ties of any kind are relatively rare unless the electorate is small. 
In very small electorates ties are more common, but
cases where Forwards, Backwards and Ordered Tie-Breaking all fail to break the
tie are unusual even then, so election by random draw will be a rare event. 

In the earliest version of the software \citep{Silverman2002}, 
ties were broken deterministically: 
if a candidate was to be elected, the last-named member of a tie was chosen. On the other hand, if there was a tie for elimination, it was the first named who was eliminated. These choices were aimed at compensating in a small way for the tendency of candidates higher up the ballot paper to get more votes.
However, they depended on position on the ballot paper, which might be 
viewed as somewhat arbitrary,
and in the \pkg{vote} package we have used a more systematic criterion.

\subsection{Equal Preference STV}
\label{subsec:equalpref}
Extant implementations of STV require that voters not give equal preferences
(except among the candidates that they do not rank). 
However, \citet{Meek1970} has pointed out that the single transferable
vote system does not exclude this possibility, and outlined how the votes
might be counted. This has never been implemented before in software,
to our knowledge, although it is used for the election of the Trustees
of the John Muir Trust \cite{WikipediaSTV2020}.

The basic idea is that if, for example, a voter gives their first preference
to candidates A and B, then the vote will be equally split between the two,
giving half a vote to each.
If A is elected, then the proportion of the half-vote for A corresponding
to A's surplus will be transferred to their next highest preference.
This will be B if B is still in contention, i.e. if B has not been elected or
eliminated by that stage. Otherwise, it will be the remaining candidate with the next highest preference from that voter.
Similarly, if A is eliminated, the half-vote for A will be fully transferred
to their next highest remaining preference. This will be B if B is still in 
contention, or otherwise the candidate with the next highest preference. 

The same principle applies if there are three or more equal preferences.
For example, consider the case where there are three equal preferences A, B and C, and A is eliminated/elected.  If A is elected, the proportion of the one-third vote for A corresponding to A's surplus is equally divided between B and C.
If A is eliminated, then both B and C get increased to a half vote.
Algebraically, this is implemented by the change below in Line 11 of 
Algorithm \ref{alg:stv}.

Otherwise the count proceeds in the same way as when equal preferences 
are not allowed. The argument \code{equal.ranking} in the \code{stv}
function is set to \code{TRUE} when equal preferences are allowed.
In this case, votes are postprocessed before counting so that they 
correctly reflect preferences. For example, a vote 1, 1, 2, 3, 3, 3
would be recoded to 1, 1, 3, 4, 4, 4. This is in contrast with the usual case 
where equal preferences are not allowed and \code{equal.ranking=FALSE}, 
when votes with non-sequential preferences,
such as 1, 2, 4, 5, are declared invalid and considered spoiled.

STV with equal preferences can be implemented by Algorithm \ref{alg:stv}
with only two relatively small changes, namely:
\begin{description}
\item[Line \ref{algline:u}] replaced by: $u_{i,j} \gets w_i \delta_{Y_{i,j}}(1) /\sum_{\ell=1}^M \delta_{Y_{i,\ell}}(1)  \quad \forall \,  i = 1, \dots, N, \; j = 1,\dots,M$
\item[Line \ref{algline:w}] replaced by:  $w_r \gets \sum_{j=1}^M u_{r,j} - u_{r,k} + u_{r,k} * S \quad \forall \, r \text{ where } Y_{r,k} = 1$
\end{description}

Note that if applied to votes with no equal preferences, the modified algorithm yields the same result as Algorithm \ref{alg:stv}. In such a case, the denominator in Line~\ref{algline:u} is equal to 1 for all $i$ and thus, $u$ is the same as in Algorithm \ref{alg:stv}. Similarly in Line~\ref{algline:w}, $\sum_{j=1}^M u_{r,j} - u_{r,k} = 0$ if there are no equal preferences and thus, $w$ is the same as in Algorithm \ref{alg:stv}.

We illustrate this functionality using the food election data by setting the first three votes to equal first preferences for Chocolate and Strawberries, instead of first and second preferences:
\begin{CodeInput}
R> food_election2 <- food_election
R> food_election2[c(1:3), 4] <- 1
R> stv (food_election2, equal.ranking = TRUE)
\end{CodeInput}
\begin{CodeOutput}
Results of Single transferable vote with equal preferences
==========================================================                           
Number of valid votes:   20
Number of invalid votes:  0
Number of candidates:     5
Number of seats:          2


|             |         1| 2-trans|     2| 3-trans|      3| 4-trans|       4|
|:------------|---------:|-------:|-----:|-------:|------:|-------:|-------:|
|Quota        |     6.668|        | 6.667|        |  6.667|        |   5.437|
|Oranges      |     4.000|   0.000| 4.000|       2|  6.000|   0.000|   6.000|
|Pears        |     2.000|   0.000| 2.000|      -2|       |        |        |
|Chocolate    |    10.500|  -3.832|      |        |       |        |        |
|Strawberries |     2.500|   2.372| 4.872|       0|  4.872|   0.000|   4.872|
|Sweets       |     1.000|   1.460| 2.460|       0|  2.460|  -2.460|        |
|Elected      | Chocolate|        |      |        |       |        | Oranges|
|Eliminated   |          |        | Pears|        | Sweets|        |        |

Elected: Chocolate, Oranges 
\end{CodeOutput}

Once again, Oranges is elected second, ahead of Strawberries, although the
margin of victory is smaller than before.

\subsection{Reserved Seats in STV}
\label{subsec:groups}
In addition to having a given number of seats to fill, it may be desired to elect a minimum number of candidates from a specified class or group of candidates. For example, the selection of plenary papers at a conference might wish to reserve at least two slots for students.  Or the election of a committee might wish to ensure that at least three women were elected.

We have incorporated this feature into the \code{stv} function as an option. Users can specify the number of reserved seats with the argument \code{group.mcan} and mark the members eligible for those seats in the argument \code{group.members}. 

When this requirement is present, our STV algorithm is modified as follows. 
Suppose $m$ denotes the number of seats and $g$ denotes the number of reserved seats and candidates are either {\em marked} (eligible for reserved seats) or {\em unmarked} (not eligible). Then on each count,

\begin{itemize}
\item if the leading candidate exceeds the quota they are elected, except that if $m-g$ unmarked candidates have already been elected, they are only elected if they are marked. Or,
\item if no candidate has been elected on this round, the candidate with the fewest votes is eliminated, except that if there are only $g$ marked candidates still in play (including any already elected) or if there are already $m - g$ unmarked candidates elected, the unmarked candidate with the fewest votes is eliminated (even if that number of votes is above the quota). 
\end{itemize}
We will illustrate the reserved seats feature in Section~\ref{subsec:committee}.

\section{Examples}
\label{sec:examples}
We now illustrate the different systems using three examples of 
elections. Perhaps ironically, systems are more robust with larger than 
smaller electorates, in the sense that 
their results are less sensitive to small changes in the electoral system.
We therefore start with a political election with
a relatively large electorate, continue with the election of the council
of a scientific organization with a moderate-sized electorate, 
and finally describe an election with a very small electorate.
Each of these was a multi-winner election, but we will also use them
to illustrate the single-winner electoral methods.

\subsection{Irish General Election 2002: Dublin West Constituency}
\label{subsec:DublinWest}
The Dublin West constitutency in the 2002 Irish general election had three
seats to be filled, nine candidates and just under 30,000 ranked votes. 
The dataset, called \code{dublin_west}, is included in the package.

\begin{CodeInput}
R> data (dublin_west)
R> head(dublin_west)
\end{CodeInput}
\begin{CodeOutput}
  Bonnie Burton Ryan Higgins Lenihan McDonald Morrissey Smyth Terry
1      0      4    0       3       0        0         1     5     2
2      0      0    2       0       1        4         3     0     0
3      0      0    3       0       1        0         2     0     0
4      0      2    0       0       0        0         3     0     1
5      0      2    1       0       0        0         0     0     0
6      0      3    2       0       1        0         0     0     0
\end{CodeOutput}

We illustrate the single-winner methods by assuming that there is
just one seat to be filled. First the plurality method. It is necessary
to convert the dataset into a set of zeros and ones to run the
\code{plurality} function:
\begin{CodeInput}
R> dublin_west1 <- 1*(dublin_west == 1)
R> plurality (dublin_west1)
\end{CodeInput}
\begin{CodeOutput}
Results of Plurality voting
===========================                              
Number of valid votes:   29988
Number of invalid votes:     0
Number of candidates:        9
Number of seats:             1


|    |Candidate | Total| Elected |
|:---|:---------|-----:|:-------:|
|1   |Lenihan   |  8086|    x    |
|2   |Higgins   |  6442|         |
|3   |Burton    |  3810|         |
|4   |Terry     |  3694|         |
|5   |McDonald  |  2404|         |
|6   |Morrissey |  2370|         |
|7   |Ryan      |  2300|         |
|8   |Bonnie    |   748|         |
|9   |Smyth     |   134|         |
|Sum |          | 29988|         |

Elected: Lenihan 
\end{CodeOutput}
Lenihan was elected, although he received only 27\% of the first preference 
votes.

Here is the two-round runoff result:
\begin{CodeInput}
R> tworound.runoff (dublin_west)
\end{CodeInput}
\begin{CodeOutput}
Results of two-round-runoff voting
==================================                              
Number of valid votes:   29988
Number of invalid votes:     0
Number of candidates:        9
Number of seats:             1


|    |Candidate | Total| Percent| ROTotal| ROPercent| Elected |
|:---|:---------|-----:|-------:|-------:|---------:|:-------:|
|1   |Bonnie    |   748|     2.5|       0|       0.0|         |
|2   |Burton    |  3810|    12.7|       0|       0.0|         |
|3   |Ryan      |  2300|     7.7|       0|       0.0|         |
|4   |Higgins   |  6442|    21.5|   12457|      47.3|         |
|5   |Lenihan   |  8086|    27.0|   13900|      52.7|    x    |
|6   |McDonald  |  2404|     8.0|       0|       0.0|         |
|7   |Morrissey |  2370|     7.9|       0|       0.0|         |
|8   |Smyth     |   134|     0.4|       0|       0.0|         |
|9   |Terry     |  3694|    12.3|       0|       0.0|         |
|Sum |          | 29988|   100.0|   26357|     100.0|         |

Elected: Lenihan 
\end{CodeOutput}
Lenihan was again elected, but this time after a run-off,
as he did not get a majority on the first count.
He got an absolute majority on the second count.
This indicates a broader base of support than the plurality vote.

We now illustrate the single-winner approval voting method by assuming that 
voters ``approved'' any candidate to whom they gave their first, 
second or third preference. 
Under this assumption, voters approved 2.8 candidates on average.

\begin{CodeInput}
R> dublin_west2 <- 1*(dublin_west == 1 | dublin_west == 2 | dublin_west == 3)
R> approval (dublin_west2)
\end{CodeInput}
\begin{CodeOutput}
Results of Approval voting
==========================                              
Number of valid votes:   29988
Number of invalid votes:     0
Number of candidates:        9
Number of seats:             1


|    |Candidate | Total| Elected |
|:---|:---------|-----:|:-------:|
|1   |Lenihan   | 15253|    x    |
|2   |Higgins   | 13638|         |
|3   |Burton    | 12863|         |
|4   |Ryan      | 10014|         |
|5   |Terry     |  9810|         |
|6   |Morrissey |  9411|         |
|7   |McDonald  |  6674|         |
|8   |Bonnie    |  4936|         |
|9   |Smyth     |   636|         |
|Sum |          | 83235|         |

Elected: Lenihan 
\end{CodeOutput}

Once again, Lenihan wins.
The multi-winner approval vote method with three seats gives wins to 
Lenihan, Higgins and Burton, because they got the most votes.

The Condorcet method did have both a winner and a loser in this case:
\begin{CodeInput}
R> condorcet (dublin_west)
\end{CodeInput}
\begin{CodeOutput}
Results of Condorcet voting
===========================                              
Number of valid votes:   29988
Number of invalid votes:     0
Number of candidates:        9
Number of seats:             1
\end{CodeOutput}
\begin{scriptsize}
\begin{CodeOutput}
|          | Bonnie| Burton| Ryan| Higgins| Lenihan| McDonald| Morrissey| Smyth| Terry| Total| Winner | Loser |
|:---------|------:|------:|----:|-------:|-------:|--------:|---------:|-----:|-----:|-----:|:------:|:-----:|
|Bonnie    |      0|      0|    0|       0|       0|        0|         0|     1|     0|     1|        |       |
|Burton    |      1|      0|    1|       0|       0|        1|         1|     1|     1|     6|        |       |
|Ryan      |      1|      0|    0|       0|       0|        1|         1|     1|     0|     4|        |       |
|Higgins   |      1|      1|    1|       0|       0|        1|         1|     1|     1|     7|        |       |
|Lenihan   |      1|      1|    1|       1|       0|        1|         1|     1|     1|     8|   x    |       |
|McDonald  |      1|      0|    0|       0|       0|        0|         0|     1|     0|     2|        |       |
|Morrissey |      1|      0|    0|       0|       0|        1|         0|     1|     0|     3|        |       |
|Smyth     |      0|      0|    0|       0|       0|        0|         0|     0|     0|     0|        |   x   |
|Terry     |      1|      0|    1|       0|       0|        1|         1|     1|     0|     5|        |       |

Condorcet winner: Lenihan
Condorcet loser: Smyth
\end{CodeOutput}
\end{scriptsize}

The STV result is as follows:
\begin{CodeInput}
R> stv.dwest <- stv (dublin_west, mcan = 3, eps = 1, digits = 0)
\end{CodeInput}
\begin{CodeOutput}
Results of Single transferable vote
===================================                              
Number of valid votes:   29988
Number of invalid votes:     0
Number of candidates:        9
Number of seats:             3
\end{CodeOutput}
\begin{tiny}
\begin{CodeOutput}
|           |       1| 2-trans|     2| 3-trans|      3| 4-trans|        4| 5-trans|       5| 6-trans|         6| 7-trans|    7| 8-trans|      8|
|:----------|-------:|-------:|-----:|-------:|------:|-------:|--------:|-------:|-------:|-------:|---------:|-------:|----:|-------:|------:|
|Quota      |    7498|        |  7491|        |   7486|        |     7465|        |    7303|        |      7233|        | 7043|        |   6143|
|Bonnie     |     748|       8|   756|      20|    776|    -776|         |        |        |        |          |        |     |        |       |
|Burton     |    3810|      55|  3865|       4|   3869|     207|     4076|     295|    4372|     211|      4583|     763| 5345|    1191|   6536|
|Ryan       |    2300|     298|  2598|      23|   2621|      65|     2686|     357|    3042|      77|      3119|     673| 3792|   -3792|       |
|Higgins    |    6442|      68|  6510|      21|   6531|     198|     6728|    1124|    7853|    -550|          |        |     |        |       |
|Lenihan    |    8086|    -588|      |        |       |        |         |        |        |        |          |        |     |        |       |
|McDonald   |    2404|      24|  2428|      19|   2447|      76|     2523|   -2523|        |        |          |        |     |        |       |
|Morrissey  |    2370|      70|  2440|      13|   2453|      98|     2551|     108|    2659|      52|      2711|   -2711|     |        |       |
|Smyth      |     134|       1|   135|    -135|       |        |         |        |        |        |          |        |     |        |       |
|Terry      |    3694|      43|  3737|      21|   3758|      69|     3828|     151|    3979|      71|      4050|     896| 4946|     802|   5748|
|Elected    | Lenihan|        |      |        |       |        |         |        | Higgins|        |          |        |     |        | Burton|
|Eliminated |        |        | Smyth|        | Bonnie|        | McDonald|        |        |        | Morrissey|        | Ryan|        |       |

Elected: Lenihan, Higgins, Burton 
\end{CodeOutput}
\end{tiny}

The three candidates elected were also the ones who got the most first
preference votes. 
All the candidates represented different political parties or were independents, except Ryan and Lenihan, who were both candidates for the Fianna F\'{a}il party, the largest party in Ireland at the time.
Lenihan was elected on the first count with a surplus of 588 votes,
and 298 of these were transferred to Ryan, the most of any candidate.
This reflects the fact that voters tend to give their highest preferences
to candidates of the same party, although here we can see that many of
the Lenihan voters did not in fact give their second preferences to Ryan.

Although this was an election with almost 30,000 votes
and the electoral system appears somewhat complex, 
the counting takes just two seconds on a Macbook Pro laptop. 

Note that the results were slightly different from the results using 
the Irish STV system, although the same candidates were elected;
see \citet{WikipediaDublinWest2020}.
This is because of several minor differences between the Irish STV system
and the \code{stv} function in the \pkg{vote} package.
The most important of these is that in the \pkg{vote} package, the quota
declines as the counts proceed, to reflect votes that are not transferred
because voters did not express enough preferences. In the Irish STV system,
the quota remains the same throughout the counts. We chose to make the 
quota adaptive because it allows a more complete transfer of the votes of 
candidates elected. However, if the argument \code{constant.quota} is set to \code{TRUE}, the quota is kept constant for all counts.

\begin{figure}[!htb]
\begin{center}
\includegraphics[height=0.45\textheight]{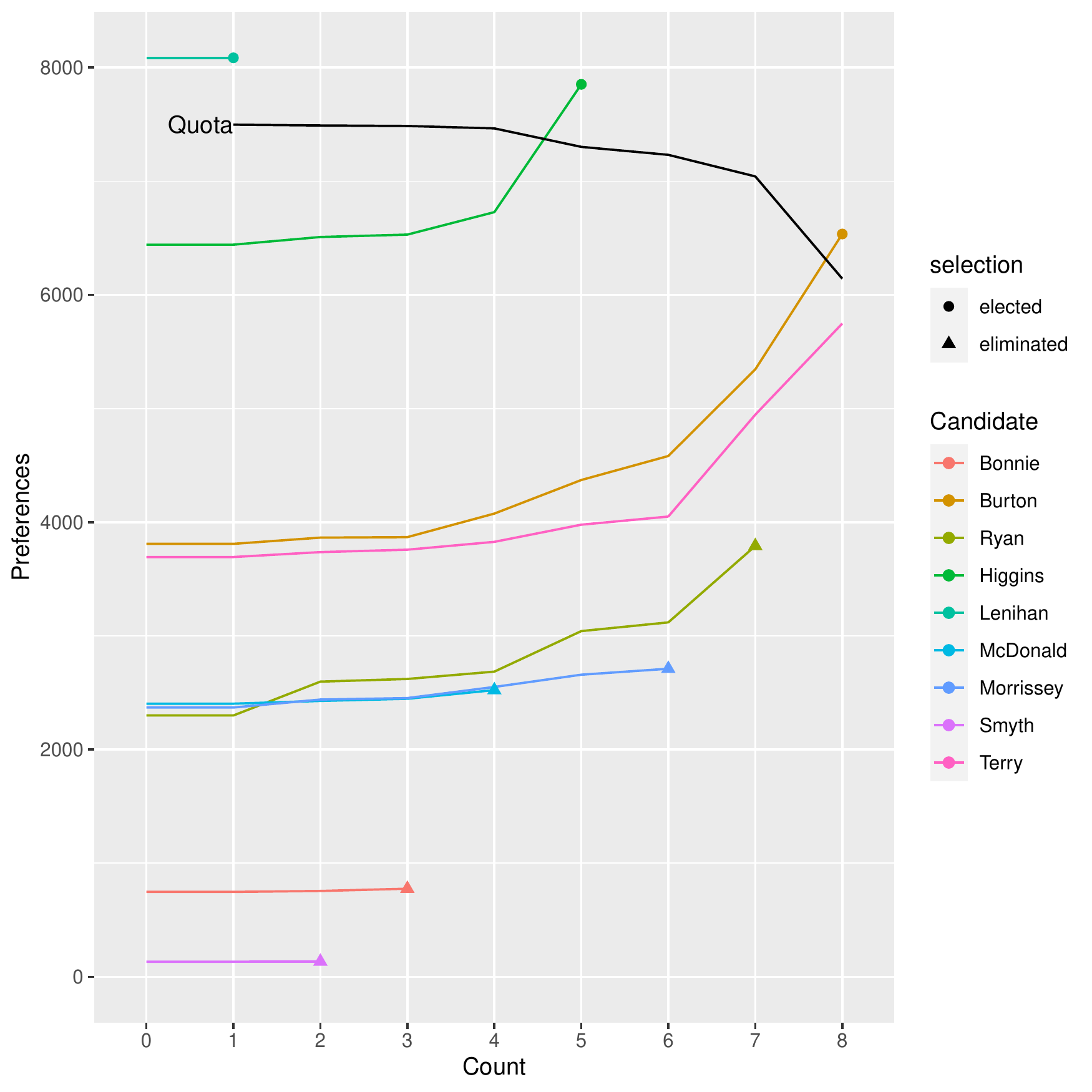}
\end{center}
\caption{\label{fig-STVdwest1} \small Evolution of candidates' votes over STV
counts in the 2002 Irish general election in Dublin West.}
\end{figure}

The STV results can be visualized in several ways. 
Figure \ref{fig-STVdwest1} has been produced by the command 
\begin{CodeInput}
R> plot (stv.dwest)
\end{CodeInput}
It shows the evolution of the candidate's vote totals over successive counts, 
as well as of the quota. It can be seen that, while candidates mostly stayed
in the same order, the candidate Ryan overtook two other candidates
thanks to transfers, even though she was eventually eliminated.
This reflects the fact that she had high preferences among voters who gave
their first preferences to Lenihan and Morrissey.

\begin{figure}[!htb]
\begin{center}
\includegraphics[height=0.5\textheight, keepaspectratio=true]{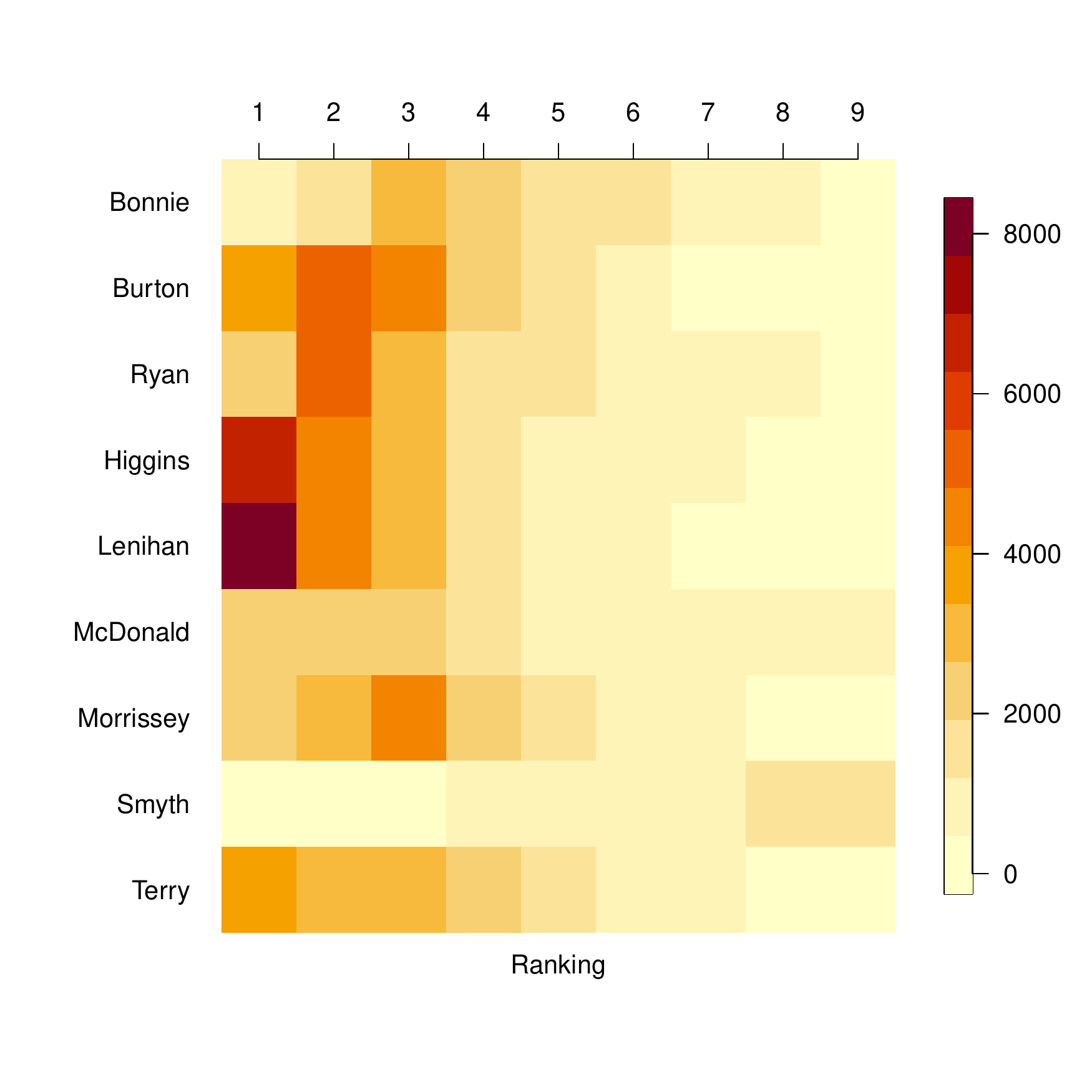}
\end{center}
\caption{\label{fig-STVdwest2} \small Number of each preference votes that each 
candidate received in the 2002 Irish general election in Dublin West.}
\end{figure}

Figure \ref{fig-STVdwest2} shows the number of each preference votes that
each candidate received. The first preferences reflect the numbers we
know from the first count. It can be seen that Ryan and Burton had the
most second preferences; in Ryan's case this is because she was the 
second Fianna F\'ail (FF) candidate behind Lenihan, and got the majority of his
second preferences. Burton and Morrissey had the most third preferences.

\begin{figure}[htb]
\begin{center}
\includegraphics[height=0.5\textheight, keepaspectratio=true]{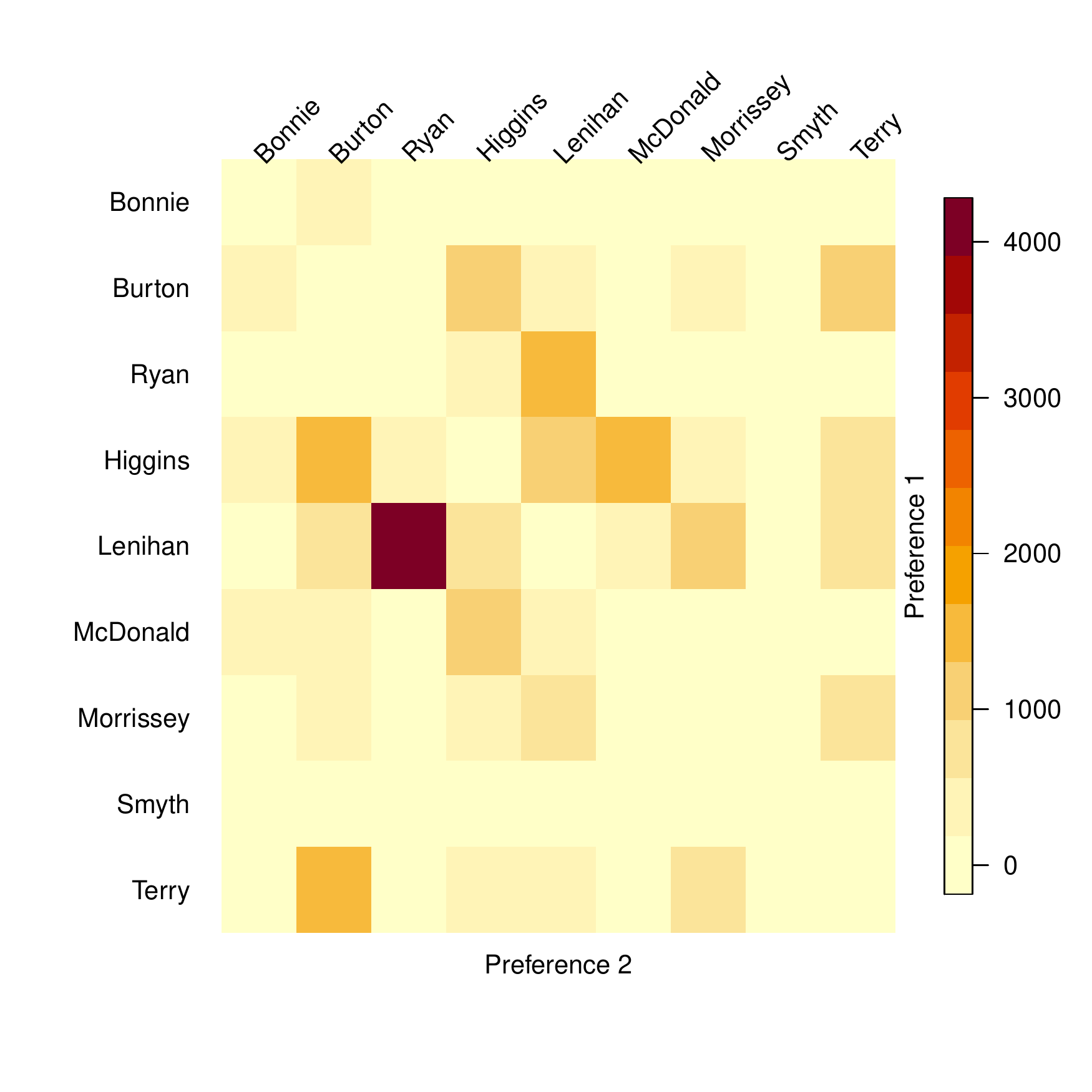}
\end{center}
\caption{\label{fig-STVdwest3} \small Number of votes for each combination of 
first and second preferences
in the 2002 Irish general election in Dublin West.}
\end{figure}

Figure \ref{fig-STVdwest3} shows the number of votes for each combination of
first and second preference. The biggest number is those who voted first for
Lenihan and then for Ryan, again reflecting that they are from the same
party, and that Lenihan had the most first preferences. 

\begin{figure}[htb]
\begin{center}
\includegraphics[height=0.5\textheight, keepaspectratio=true]{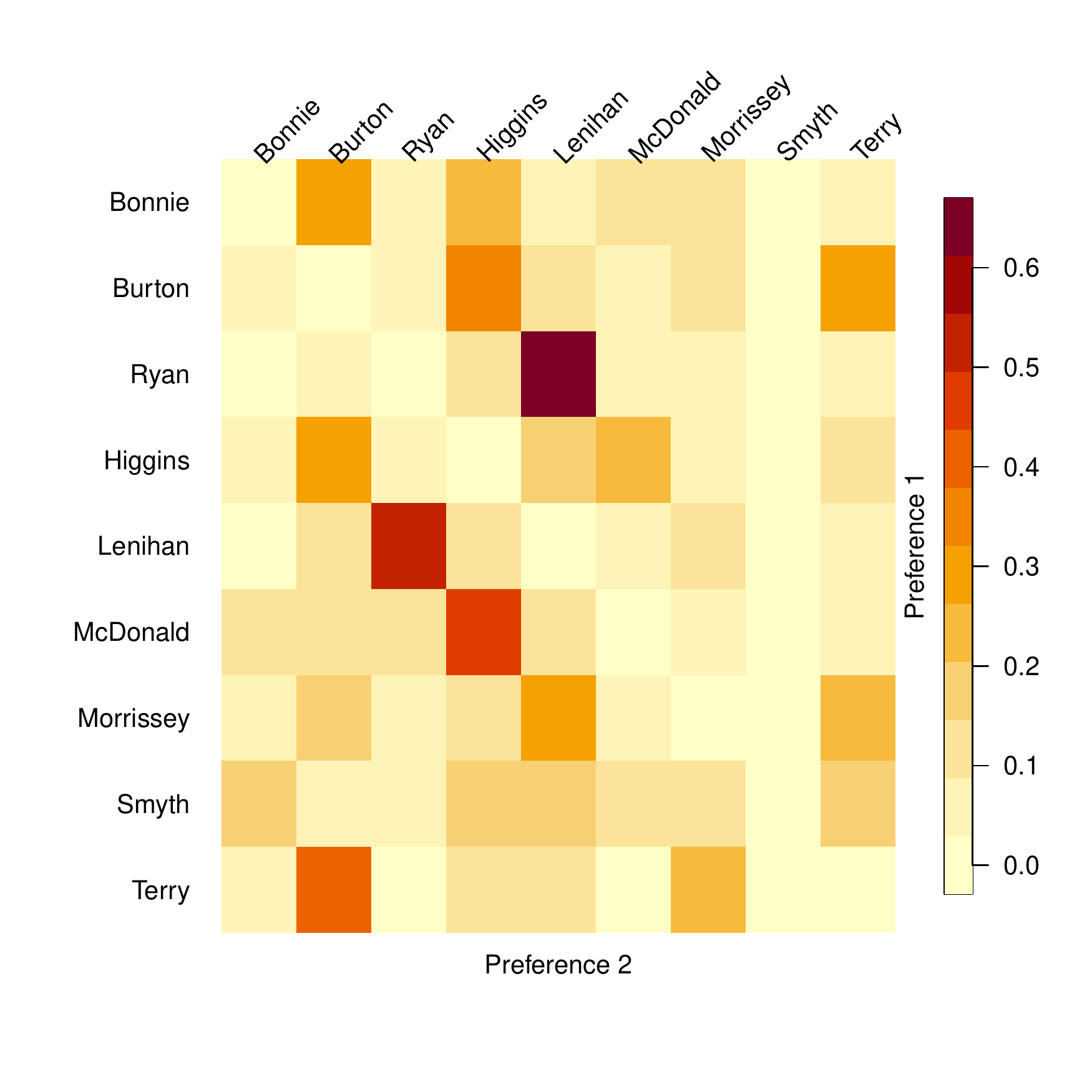}
\end{center}
\caption{\label{fig-STVdwest4} \small Proportion of the first preference votes
for each candidate that gave their second preference vote to each other
candidate  in the 2002 Irish general election in Dublin West.}
\end{figure}

Figure \ref{fig-STVdwest4} shows the same information, but in the 
form of the {\it proportion} of the first preference voters for each candidate
that cast their second preference votes for each other candidate.
The largest single cell shows that over 60\% of Ryan voters cast their
second preferences for Lenihan. 

The code for producing Figures \ref{fig-STVdwest2}, \ref{fig-STVdwest3} 
and \ref{fig-STVdwest4} is as follows:
\begin{CodeInput}
R> image (stv.dwest, all.pref = TRUE)  # Figure 2
R> image (stv.dwest, proportion = FALSE) # Figure 3
R> image (stv.dwest, proportion = TRUE) # Figure 4
\end{CodeInput}
Note that the \code{image} method is available for all functions in the package that use ranked votes, namely, in addition to \code{stv}, \code{condorcet} and \code{tworound.runoff}. However, the method cannot be used if equal preferences are present in the ballots.

\pagebreak
\subsection{IMS Council Election}
\label{subsec:IMS}
The \code{ims_election} dataset contains the votes in a past election for the
Council of the Institute of Mathematical Statistics (IMS). 
There were four seats to be filled with 10 candidates running, and 620 voters.
The names of the candidates have been anonymized\footnote{To ensure
confidentiality, the names of the candidates were replaced by arbitrarily
chosen first names that have no connection to the actual names of the 
candidates.}.
The election was carried out by STV. The results were:
\begin{CodeInput}
R> data (ims_election)
R> stv.ims <- stv (ims_election, mcan = 4, eps = 1, digits = 0)
\end{CodeInput}

\begin{center}
\includegraphics[width=\textwidth]{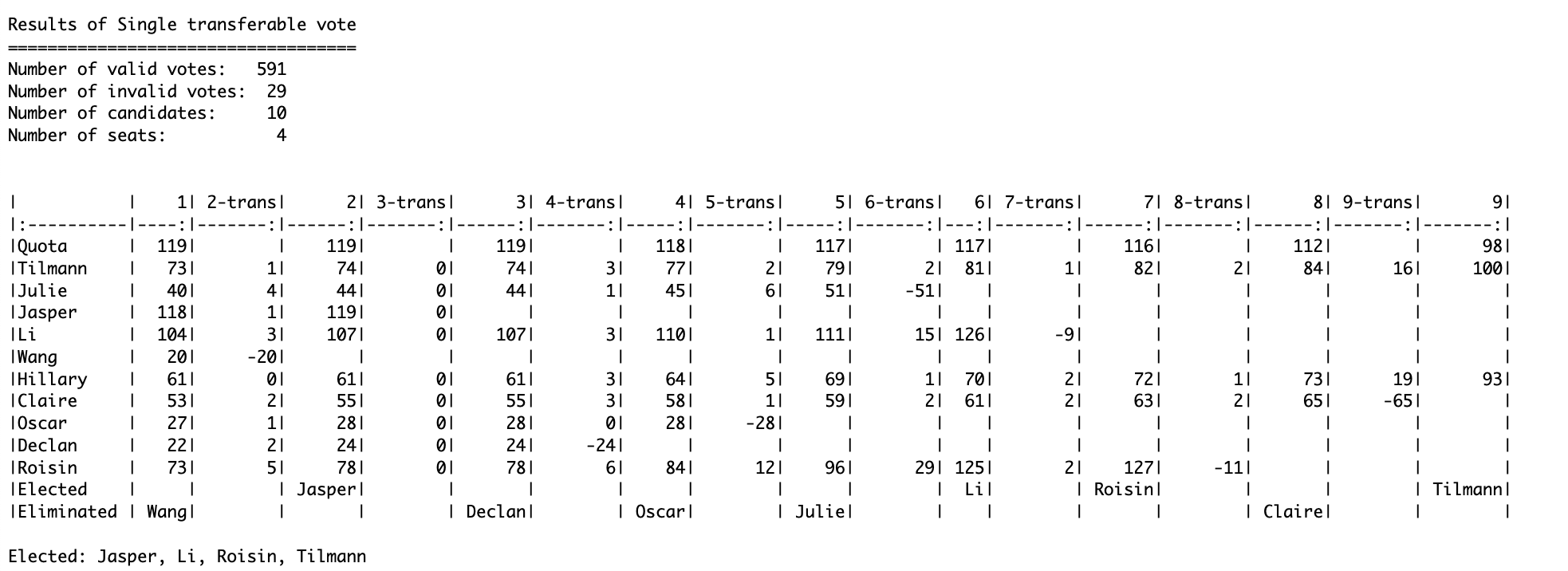}
\end{center}

\begin{figure}[!htp]
\begin{center}
\begin{tabular}{cc}
\includegraphics[width=0.5\textwidth]{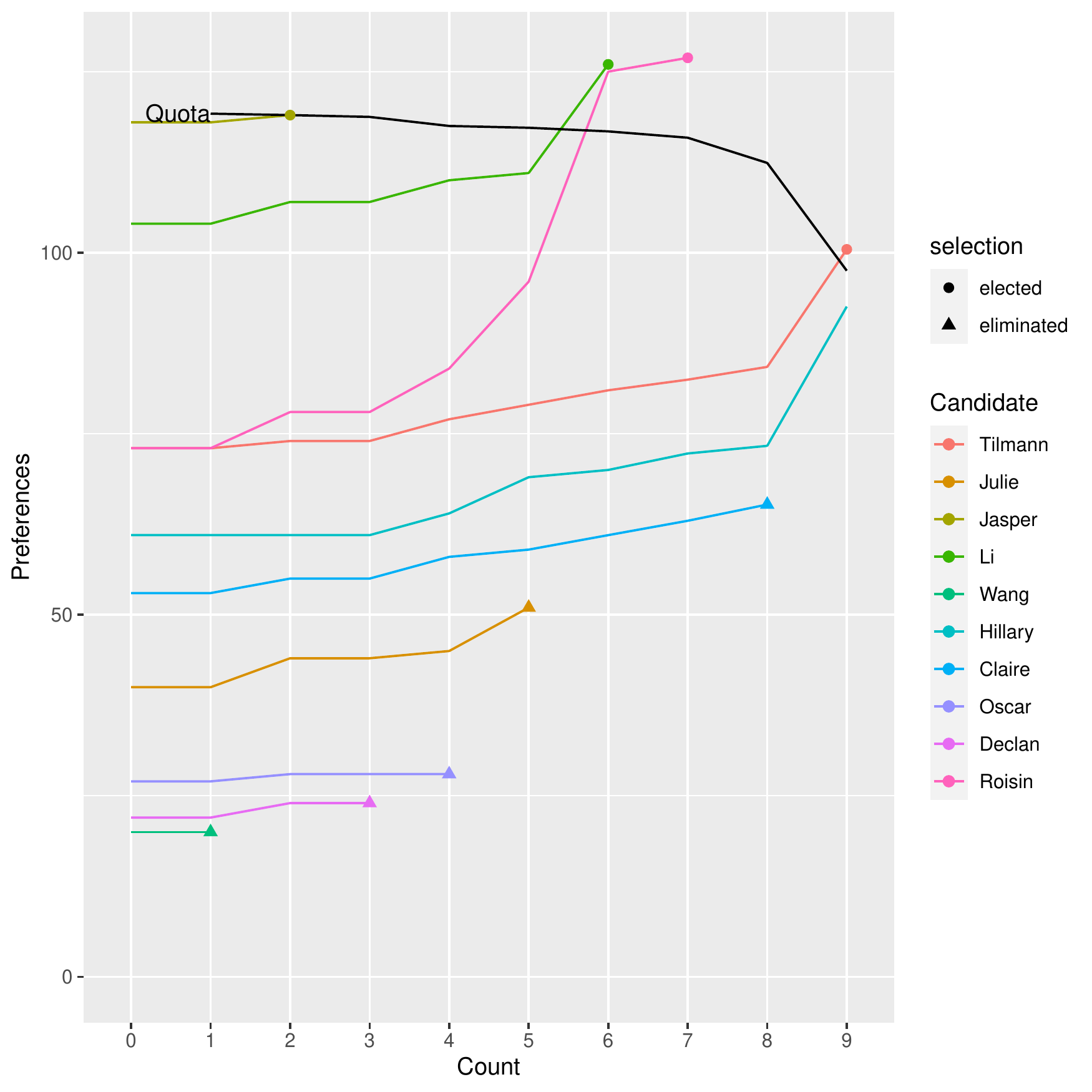} &
\includegraphics[width=0.5\textwidth]{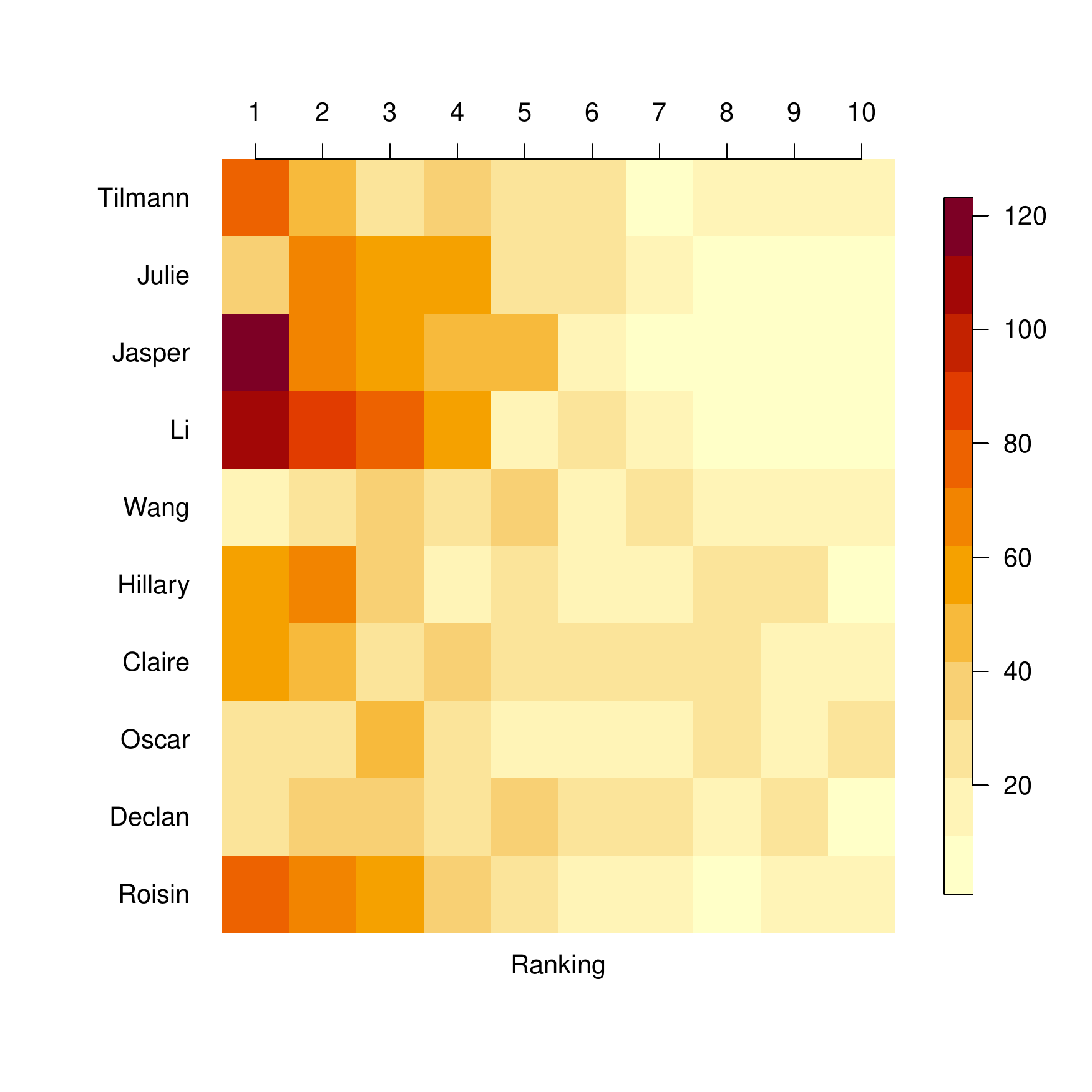} \\
\includegraphics[width=0.5\textwidth]{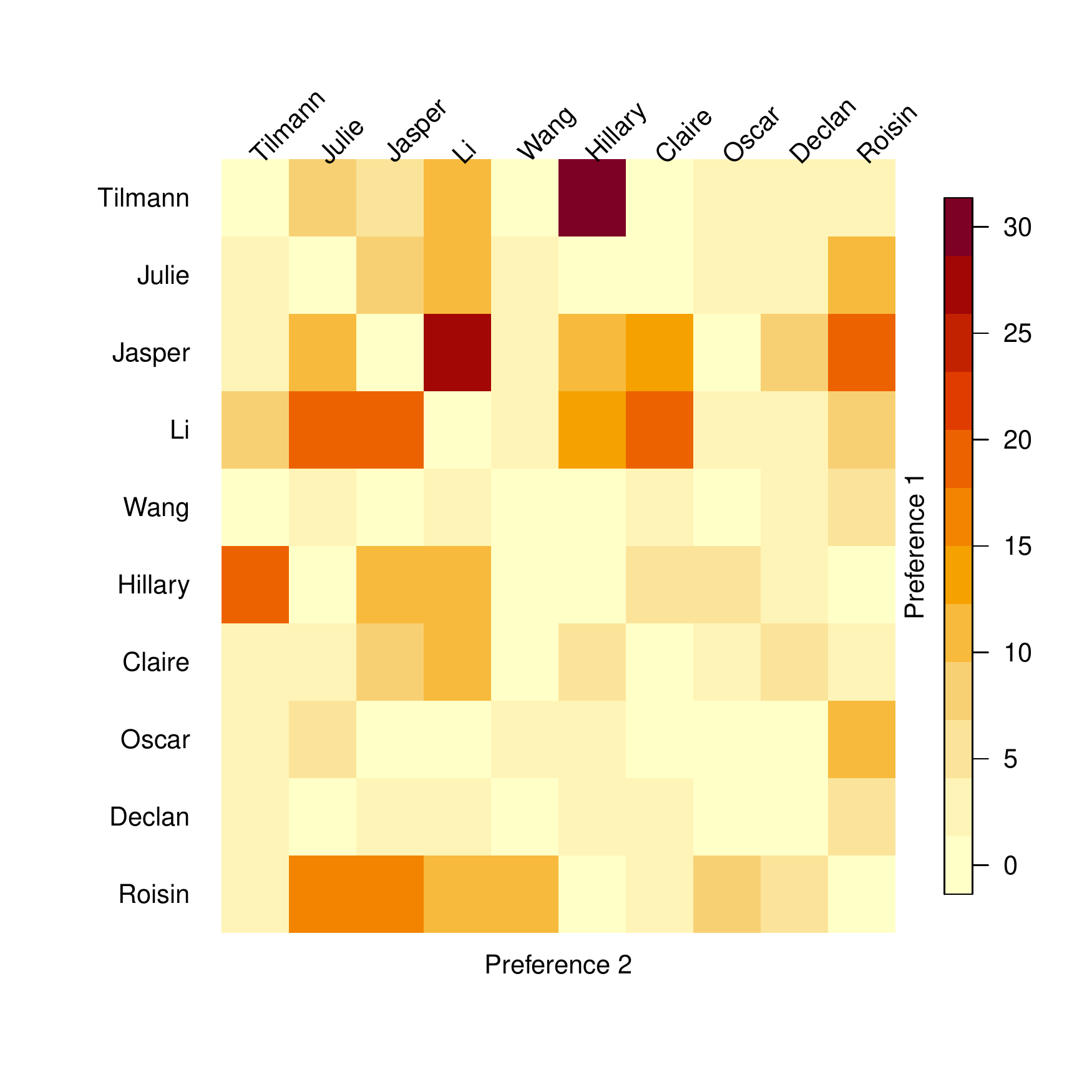} &
\includegraphics[width=0.5\textwidth]{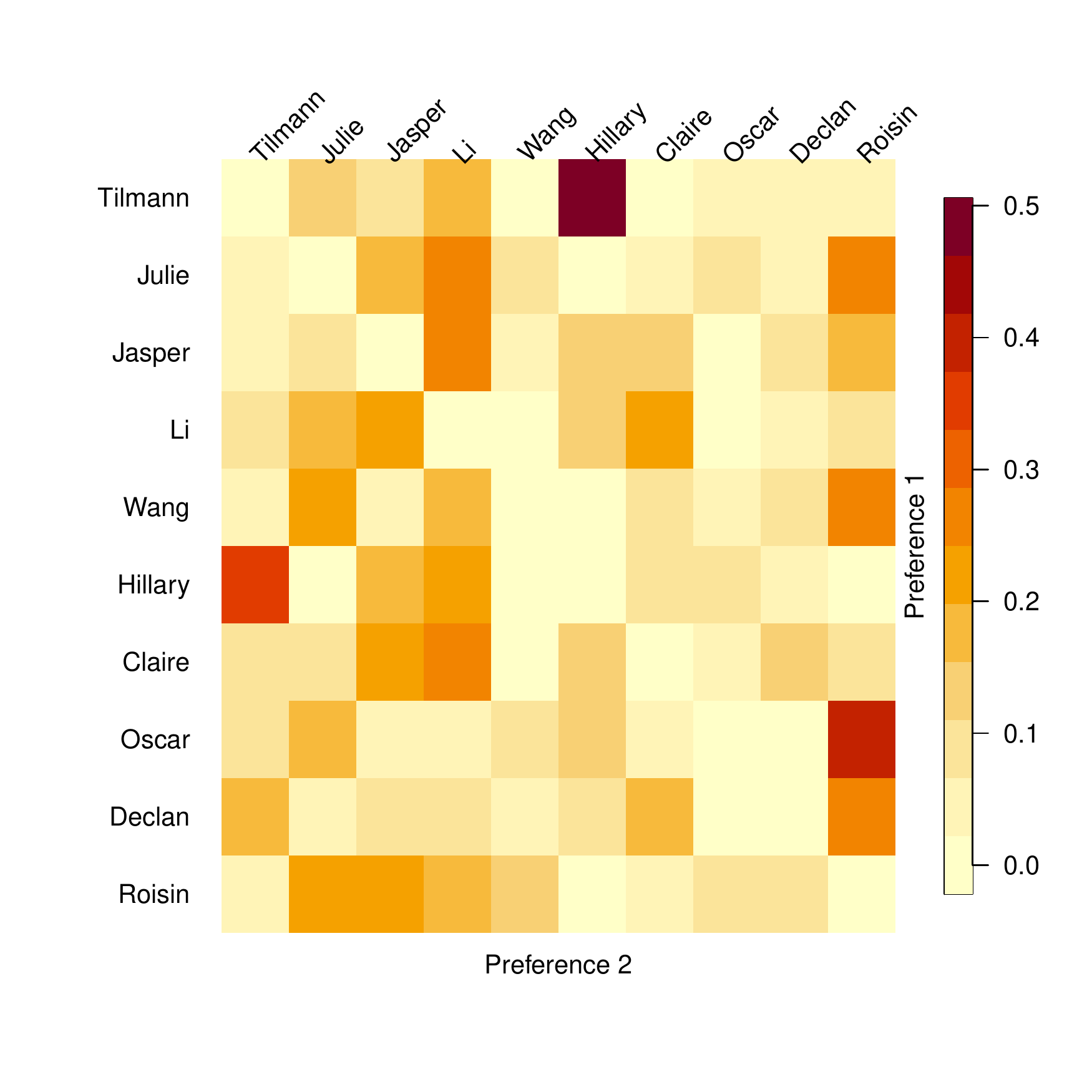}  
\end{tabular}
\end{center}
\caption{\label{fig-IMS} \small Visualization of results of IMS Council election by 
STV. (a) Top left: Evolution of votes over counts. 
(b) Top right: Number of votes for each candidate at each preference level. 
(c) Bottom left: Number of votes for each first and second  preference combination.
(d) Bottom right: Number of second  preferences as a proportion of the number of
first preference voters for each candidate.
}
\end{figure}

The results are shown in Figure \ref{fig-IMS}. 
Although the electorate was much smaller, the results show some common 
patterns to those from Dublin West. The quota declined slowly in the early
counts, and more rapidly in the later ones. The four candidates elected
were the ones that got the most first preferences. Figure \ref{fig-IMS}(d)
shows that, while there are no political parties in this election,
Tilmann and Hillary tended to share voters, as did Jasper and Li. 
We do not know the identities of the candidates because their names have been
anonymized, but 
these pairs of candidates clearly appeal to the same voters, perhaps because of
geographical or intellectual commonalities. 

However, neither Li nor Hillary was able to benefit from these shared preferences in this election. While Jasper was elected on the second count, he reached exactly the number of votes needed to reach the quota, namely 119, and thus no surplus was available for a transfer. Tilmann on the other hand was elected last, after which the election ended. If there had been one more seat available (i.e. \code{mcan = 5}), Hillary would have got Tilmann's surplus and then would have
been elected. 


\subsection{Trial Faculty Recruitment Vote}
\label{subsec:committee}
This is a trial election that was carried out to test a proposed use of STV
in a university statistics department for selecting faculty job candidates 
to whom to make offers. There were two jobs to be filled, five finalists, 
and ten voters.  
It was desired to select the two candidates to whom to make offers,
and also to produce a ranking of the other candidates. 
This is fairly typical of such elections. 
The candidates were named Augustin-Louis Cauchy, Carl Friedrich Gauss, 
Pierre-Simon Laplace, Florence Nightingale, and Sim\'{e}on Poisson.

The voters entered their choices into a web-based survey which was then converted into a text file. Here we create the corresponding dataset manually:
\begin{CodeInput}
R> faculty <- data.frame(
    Cauchy =      c(3, 4, 4, 4, 4, 5, 4, 5, 5, 5),
    Gauss  =      c(4, 1, 2, 2, 2, 2, 2, 2, 2, 4),
    Laplace =     c(5, 2, 1, 3, 1, 3, 3, 4, 4, 1),
    Nightingale = c(1, 3, 5, 1, 3, 1, 5, 1, 1, 2),
    Poisson =     c(2, 5, 3, 5, 5, 4, 1, 3, 3, 3)
    )
\end{CodeInput}

The results of the STV election were as follows:
\begin{CodeInput}
R> stv.faculty <- stv (faculty, mcan = 2, digits = 2, complete.ranking = TRUE)
\end{CodeInput}
\begin{CodeOutput}
Results of Single transferable vote
===================================                           
Number of valid votes:   10
Number of invalid votes:  0
Number of candidates:     5
Number of seats:          2

|            |           1| 2-trans|      2| 3-trans|       3| 4-trans|     4|
|:-----------|-----------:|-------:|------:|-------:|-------:|-------:|-----:|
|Quota       |        3.33|        |   3.33|        |    3.33|        |  3.33|
|Cauchy      |        0.00|    0.00|   0.00|       0|        |        |      |
|Gauss       |        1.00|    1.33|   2.33|       0|    2.33|    1.33|  3.67|
|Laplace     |        3.00|    0.00|   3.00|       0|    3.00|    0.00|  3.00|
|Nightingale |        5.00|   -1.67|       |        |        |        |      |
|Poisson     |        1.00|    0.33|   1.33|       0|    1.33|   -1.33|      |
|Elected     | Nightingale|        |       |        |        |        | Gauss|
|Eliminated  |            |        | Cauchy|        | Poisson|        |      |

Complete Ranking
================

| Rank|Candidate   | Elected |
|----:|:-----------|:-------:|
|    1|Nightingale |    x    |
|    2|Gauss       |    x    |
|    3|Laplace     |         |
|    4|Poisson     |         |
|    5|Cauchy      |         |

Elected: Nightingale, Gauss 
\end{CodeOutput}

Nightingale and Gauss were elected. The complete ranking could be useful for a vote like this, where an ordering
beyond the winning candidates may be desired, for example to make further
offers if one of the top two declines the offer.
Note that the complete ranking is conditional on the 
pre-specified number of seats or winners in the election.

\begin{figure}[!htp]
\begin{center}
\begin{tabular}{cc}
\includegraphics[width=0.5\textwidth]{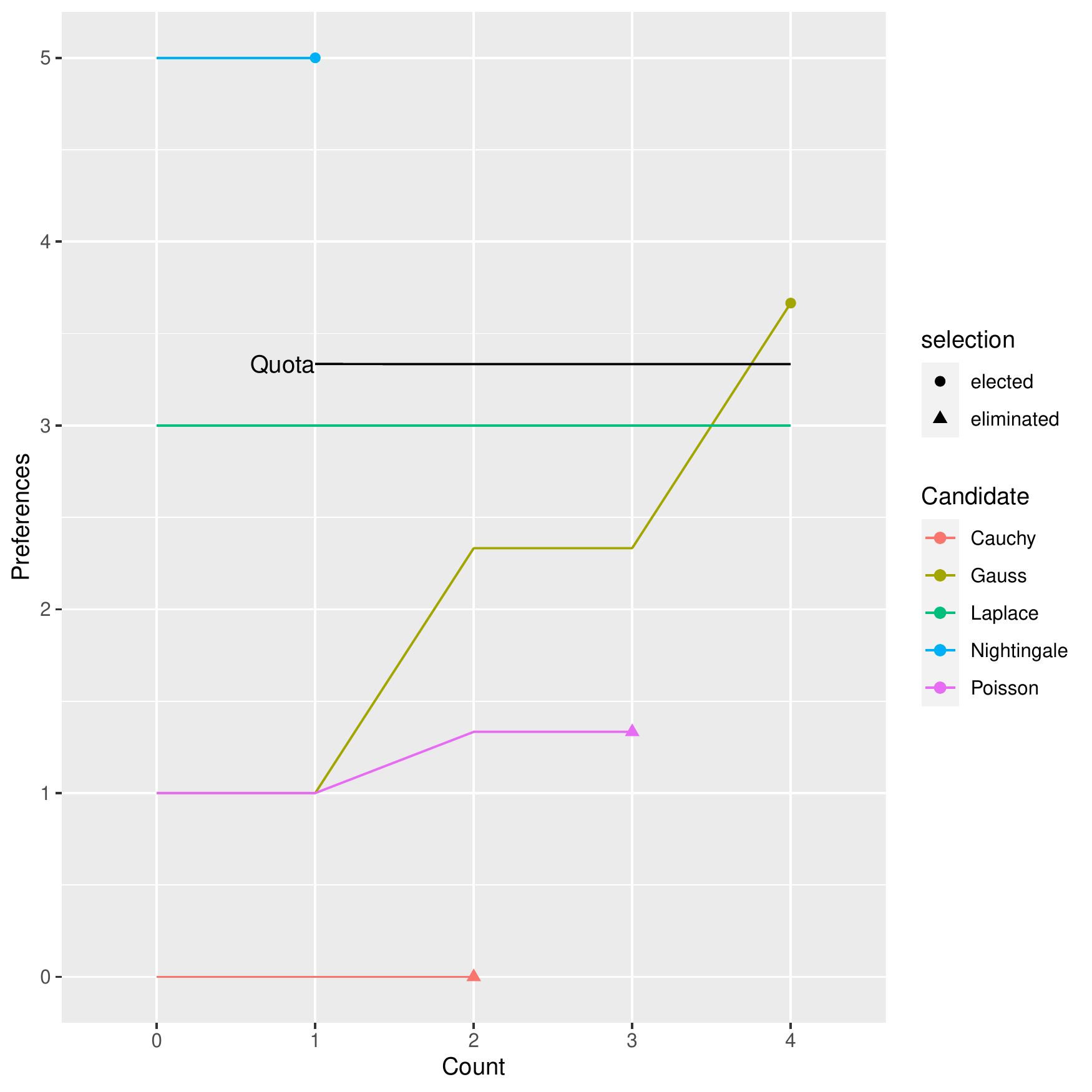} &
\includegraphics[width=0.5\textwidth]{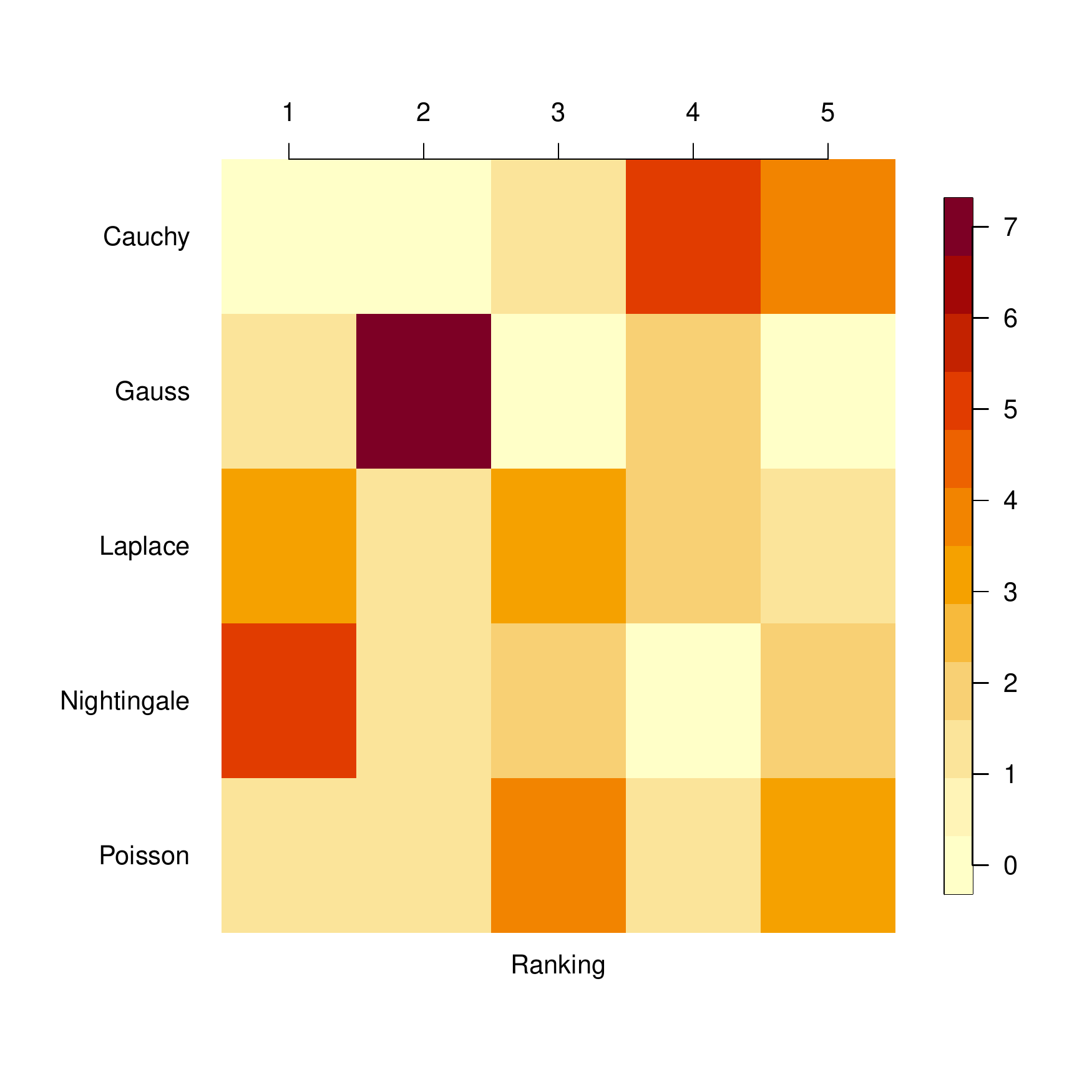} \\
\includegraphics[width=0.5\textwidth]{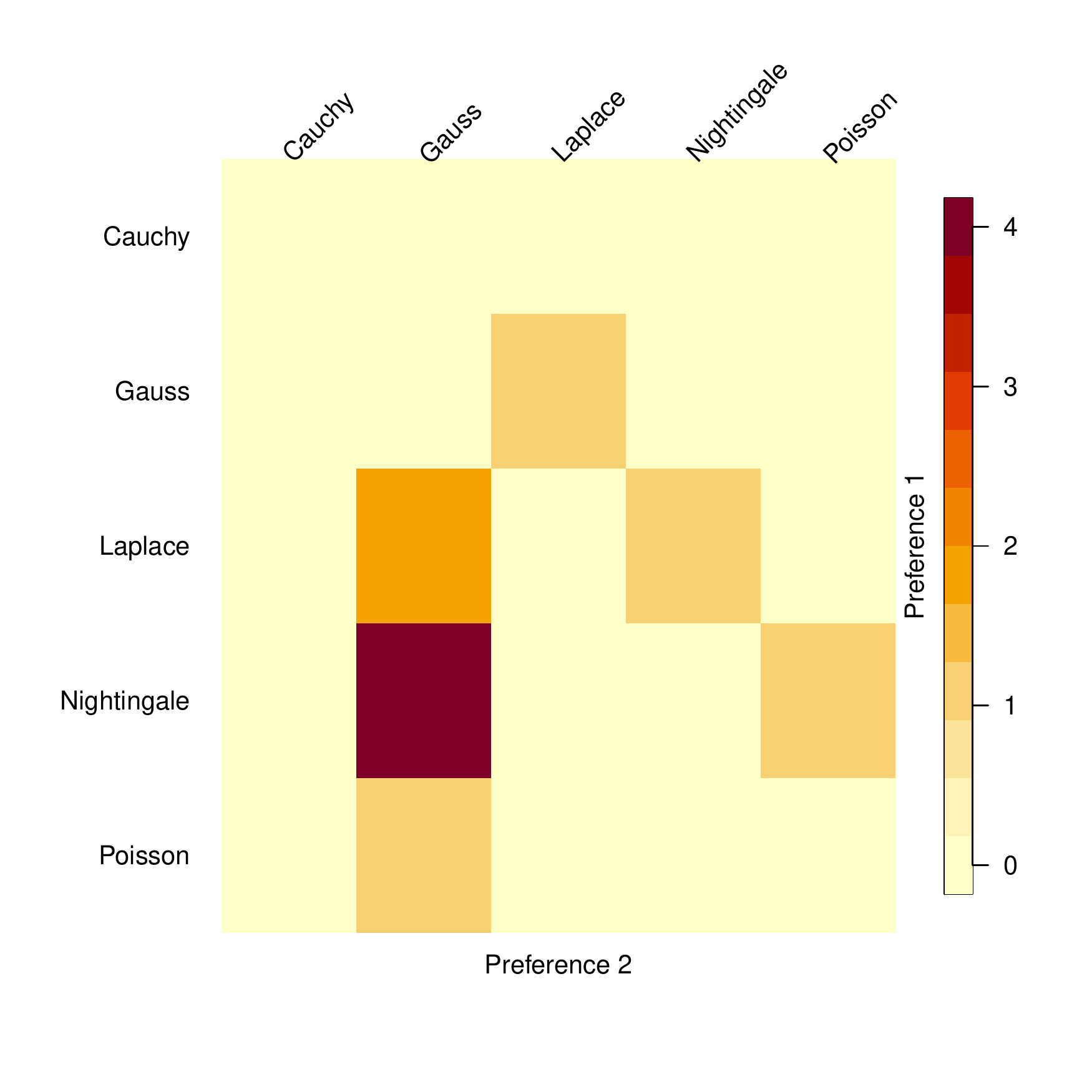} &
\includegraphics[width=0.5\textwidth]{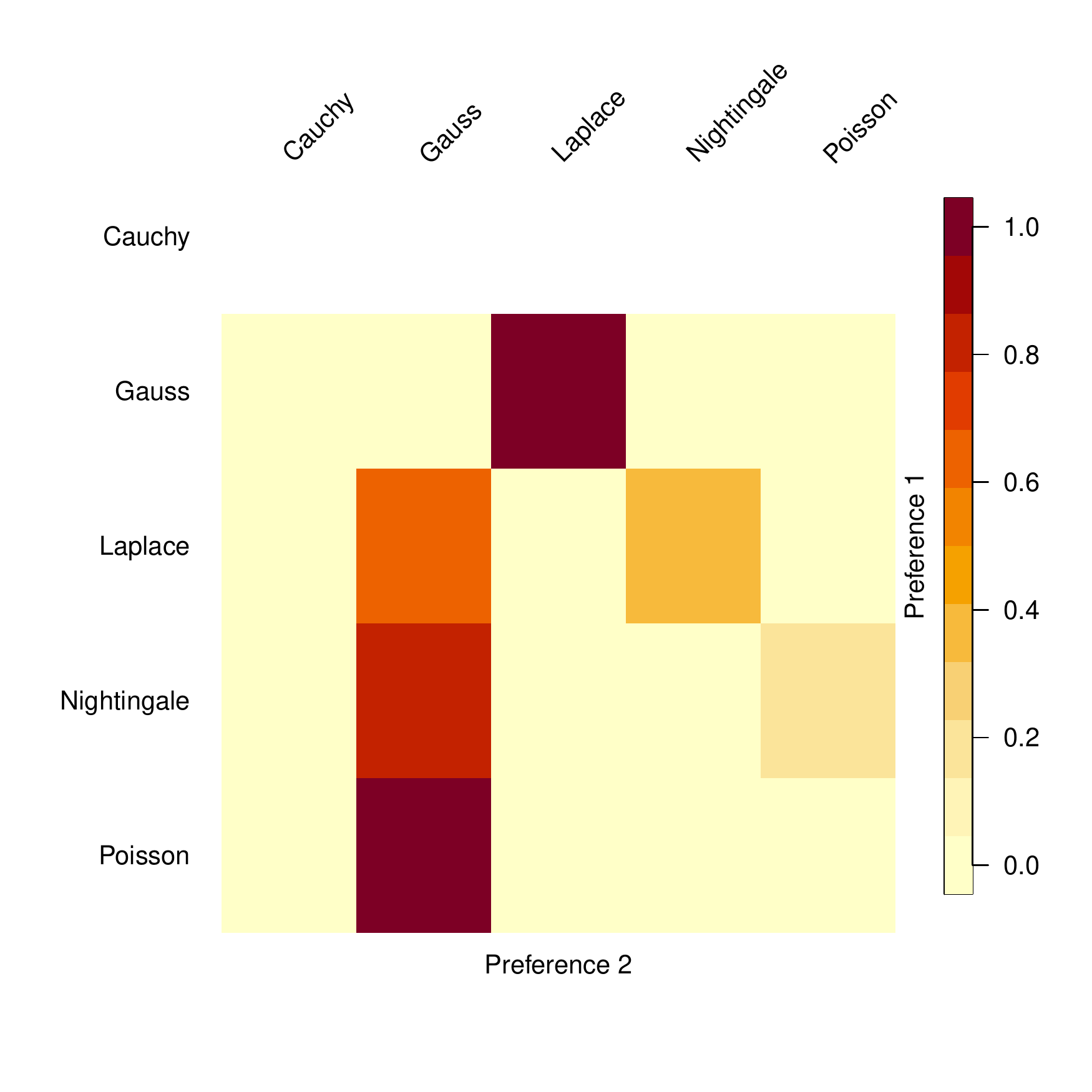}  
\end{tabular}
\end{center}
\caption{\label{fig-faculty} \small Visualization of results of the trial faculty
recruitment vote by STV. (a) Top left: Evolution of votes over counts. 
(b) Top right: Number of votes for each candidate at each preference level. 
(c) Bottom left: Number of votes for each first and second  preference combination.
(d) Bottom right: Number of second  preferences as a proportion of the number of
first preference voters for each candidate.
}
\end{figure}

The results are illustrated in Figure \ref{fig-faculty}.
An interesting feature that can be seen from Figure \ref{fig-faculty}(a) is that Laplace got more first preference votes than Gauss, but Gauss ended up beating him by a small margin for the second offer because almost every voter gave Gauss either their first or second preference. Thus, as other candidates were elected or eliminated, their votes were transferred to Gauss rather than Laplace. 
The large number of second preferences for Gauss is apparent from
Figure \ref{fig-faculty}(b). Figure \ref{fig-faculty}(c) and especially
Figure \ref{fig-faculty}(d) show that Gauss got the highest number and 
proportion of second preference votes from the electors of each of the other
candidates.

If this had been done by approval voting, and all the voters had approved their top two choices, the same two candidates would have been selected as by STV (i.e. Nightingale and Gauss). 

It is interesting to note that there was no Condorcet winner in this election,
even though Nightingale was far ahead of the other candidates by most 
criteria:
\begin{CodeInput}
> condorcet (faculty)
\end{CodeInput}
\begin{CodeOutput}
Results of Condorcet voting
===========================                           
Number of valid votes:   10
Number of invalid votes:  0
Number of candidates:     5
Number of seats:          0

|            | Cauchy| Gauss| Laplace| Nightingale| Poisson| Total| Loser |
|:-----------|------:|-----:|-------:|-----------:|-------:|-----:|:-----:|
|Cauchy      |      0|     0|       0|           0|       0|     0|   x   |
|Gauss       |      1|     0|       1|           0|       1|     3|       |
|Laplace     |      1|     0|       0|           0|       1|     2|       |
|Nightingale |      1|     1|       0|           0|       1|     3|       |
|Poisson     |      1|     0|       0|           0|       0|     1|       |

There is no condorcet winner (no candidate won over all other candidates).
Condorcet loser: Cauchy
\end{CodeOutput}
This illustrates the fact that even in a relatively clearcut case 
there may be no Condorcet winner.

To illustrate the feature of reserved seats in STV, let us assume that it is required that at least one French candidate be selected. Then,
\begin{CodeInput}
R> stv (faculty, mcan = 2, group.mcan = 1, 
+     group.members = c("Laplace", "Poisson", "Cauchy"), digits = 2)
\end{CodeInput}
\begin{CodeOutput}
Results of Single transferable vote
===================================                           
Number of valid votes:   10
Number of invalid votes:  0
Number of candidates:     5
Number of seats:          2
Number of reserved seats:          1 
Eligible for reserved seats:       3 


|            |           1| 2-trans|     2| 3-trans|       3|
|:-----------|-----------:|-------:|-----:|-------:|-------:|
|Quota       |        3.33|        |  3.33|        |    3.33|
|Cauchy*     |        0.00|    0.00|  0.00|    0.00|    0.00|
|Gauss       |        1.00|    1.33|  2.33|   -2.33|        |
|Laplace*    |        3.00|    0.00|  3.00|    1.67|    4.67|
|Nightingale |        5.00|   -1.67|      |        |        |
|Poisson*    |        1.00|    0.33|  1.33|    0.67|    2.00|
|Elected     | Nightingale|        |      |        | Laplace|
|Eliminated  |            |        | Gauss|        |        |

Elected: Nightingale, Laplace 
\end{CodeOutput}
Here, the modifications to the algorithm described in Section~\ref{subsec:groups} ensured that none of the French candidates was eliminated on the second count, as the only seat left at that point was the reserved seat. Thus, Gauss, the only non-French candidate left, was eliminated in spite of having more votes than Cauchy or Poisson. Laplace was then elected on the following count. In the output, the candidates eligible for reserved seats are marked with a star.

We now modify this dataset slightly to illustrate the equal ranking
STV method. Four of the votes were changed so as to include equal preferences:
\begin{CodeInput}
R> faculty2 <- faculty
R> faculty2[1,] <- c(2,2,3,1,1)
R> faculty2[4,] <- c(3,1,2,1,3)
R> faculty2[9,] <- c(4,1,3,1,2)
R> faculty2[10,] <- c(2,1,1,1,1)
R> faculty2
\end{CodeInput}
\begin{CodeOutput}
   Cauchy Gauss Laplace Nightingale Poisson
1       2     2       3           1       1
2       4     1       2           3       5
3       4     2       1           5       3
4       3     1       2           1       3
5       4     2       1           3       5
6       5     2       3           1       4
7       4     2       3           5       1
8       5     2       4           1       3
9       4     1       3           1       2
10      2     1       1           1       1
\end{CodeOutput}

The results of the STV election with equal preferences were as follows:
\begin{CodeInput}
R> stv.faculty.equal <- stv (faculty2, equal.ranking = TRUE, digits = 2)
\end{CodeInput}
\begin{footnotesize}
\begin{CodeOutput}
Results of Single transferable vote with equal preferences
==========================================================                           
Number of valid votes:   10
Number of invalid votes:  0
Number of candidates:     5
Number of seats:          2

|            |           1| 2-trans|      2| 3-trans|       3| 4-trans|     4|
|:-----------|-----------:|-------:|------:|-------:|-------:|-------:|-----:|
|Quota       |        3.33|        |   3.33|        |    3.33|        |  3.33|
|Cauchy      |        0.00|    0.00|   0.00|       0|        |        |      |
|Gauss       |        2.25|    0.34|   2.59|       0|    2.59|    1.69|  4.28|
|Laplace     |        2.25|    0.01|   2.26|       0|    2.26|    0.13|  2.39|
|Nightingale |        3.75|   -0.42|       |        |        |        |      |
|Poisson     |        1.75|    0.06|   1.81|       0|    1.81|   -1.81|      |
|Elected     | Nightingale|        |       |        |        |        | Gauss|
|Eliminated  |            |        | Cauchy|        | Poisson|        |      |

Elected: Nightingale, Gauss 

Warning message:
In correct.ranking(votes, quiet = quiet) :
  Votes 1, 4, 9, 10 were corrected to comply with the required format.
\end{CodeOutput}
\end{footnotesize}

The warning message indicates that the ranking was corrected.
When \code{equal.ranking=TRUE}, this correction will be made with any input, 
as long as the preferences are recorded as positive numbers (not necessarily integers). 
The corrected votes are stored in the \code{data} element of the resulting object:
\begin{CodeInput}
R> stv.faculty.equal$data[c(1, 4, 9, 10),]
\end{CodeInput}
\begin{CodeOutput}
   Cauchy Gauss Laplace Nightingale Poisson
1       3     3       5           1       1
4       4     1       3           1       4
9       5     1       4           1       3
10      5     1       1           1       1
\end{CodeOutput}
Such a correction is not made in the default case in which \code{equal.ranking=FALSE}, when the
preferences have to be an ordered sequence of integers starting at one,
with no ties and no gaps. However, votes can be corrected in the same way also from outside \code{stv}, using the function \code{correct.ranking}.

Finally, we give the results when there is a single winner to illustrate 
tie-breaking, as it so happens that tie-breaking is needed 
on two different counts in this case:
\begin{CodeInput}
R> stv.faculty.tie <- stv (faculty, mcan = 1)
\end{CodeInput}
\begin{footnotesize}
\begin{CodeOutput}
Results of Single transferable vote
===================================                           
Number of valid votes:   10
Number of invalid votes:  0
Number of candidates:     5
Number of seats:          1

|            |      1| 2-trans|       2| 3-trans|     3| 4-trans|       4| 5-trans|           5|
|:-----------|------:|-------:|-------:|-------:|-----:|-------:|-------:|-------:|-----------:|
|Quota       |  5.001|        |   5.001|        | 5.001|        |   5.001|        |       5.001|
|Cauchy      |  0.000|       0|        |        |      |        |        |        |            |
|Gauss       |  1.000|       0|   1.000|       1| 2.000|      -2|        |        |            |
|Laplace     |  3.000|       0|   3.000|       0| 3.000|       2|   5.000|      -5|            |
|Nightingale |  5.000|       0|   5.000|       0| 5.000|       0|   5.000|       5|      10.000|
|Poisson     |  1.000|       0|   1.000|      -1|      |        |        |        |            |
|Tie-breaks  |       |        |      fo|        |      |        |       f|        |            |
|Elected     |       |        |        |        |      |        |        |        | Nightingale|
|Eliminated  | Cauchy|        | Poisson|        | Gauss|        | Laplace|        |            |

Elected: Nightingale 
\end{CodeOutput}
\end{footnotesize}

On the second count, Gauss and Poisson both had one vote, the lowest number,
and so were tied for elimination. The Forwards Tie-Breaking method did not
break the tie, as they both had the same number of votes also on the first 
count. The Ordered method did break the tie, however, because Gauss had
7 second preferences, and Poisson had only 1, so Poisson was eliminated. 
The notation ``fo'' in the Tie-breaks row indicates the tie-breaking
method used, here Forwards followed be Ordered. 

On the fourth count, Laplace and Nightingale were tied with 5 votes each,
so they were tied for elimination as neither reached the quota of 5.001.
The Forwards Tie-Breaking method was
then used, and involved looking first at their numbers of votes on the 
first count, when Laplace had 3 votes and Nightingale had 5. 
As a result, Laplace was eliminated and then Nightingale was elected.
If the Backwards Tie-Breaking method had been used (by setting \code{ties = "b"}), the comparison would have been done based on the third count instead of the first count. Here too, Laplace had 3 votes and Nightingale had 5 on the third count and thus, Laplace would have been eliminated. 

Note that the ranking used by the Ordered method can be viewed via the \code{ordered.tiebreak} function:
\begin{CodeInput}
R> ordered.tiebreak(stv.faculty.tie$data)
\end{CodeInput}
\begin{CodeOutput}
     Cauchy       Gauss     Laplace Nightingale     Poisson
          1           3           4           5           2 
attr(,"sampled")
[1] FALSE FALSE FALSE FALSE FALSE
\end{CodeOutput}
It gives the elimination ranking. When used for electing a candidate, the order is reversed. The attribute ``sampled" indicates for each candidate whether sampling was involved to determine its rank, which was not the case in our example. The function \code{ordered.preferences} can be used to view the matrix of preference counts from which the ordered ranking is derived. It gives the same information as the \code{image} plot with \code{all.pref = TRUE}, but in matrix form.

\pagebreak
\section{Discussion}
\label{sec:discussion}
We have described and illustrated the \pkg{vote} package in \proglang{R},
which implements several electoral systems, namely the plurality,
two-round runoff, approval, score and single transferable vote (STV) systems
\citep{vote}.
It also identifies the Condorcet winner and loser, if they exist. It implements 
the single transferable vote system with equal preferences, the first time
this has been implemented in software to our knowledge.
It also provides several ways of visualizing the STV results.

We are not advocating any electoral system, and indeed it is well known
that no one system satisfies all of a set of criteria that one might
reasonably want to hold. Thus which system one uses can depend
on the purpose of the election. However, we are particularly interested
in multi-winner elections with small electorates, such as committee and 
council elections in organizations, and the selection of multiple job candidates, award winners or other choices by small ``selectorates.''
Such elections are common and there is no universally accepted method
for conducting them. We have found the STV system to work well in practice
for such elections, and so we have emphasized it here, giving several examples.

For completeness, we note that the most widely used political voting system around the world is a party list approach, where voters vote for a party rather than for individuals, and some mechanism is then used to fill the party slots allocated \citep{ERS2020}. 
Such systems are not relevant for the purposes of our primary interest.  

There are several other \proglang{R} packages that implement electoral systems.
The \pkg{votesys} package implements several electoral methods, including
several that are not included in the \pkg{vote} package \citep{votesys}.
It implements the instant runoff system (IRV), which is the special case
of STV for single-winner elections, but it does not implement the full version
of STV for multi-winner elections. The \pkg{rcv} package also implements
IRV (calling it Ranked Choice Voting), but has been removed from CRAN
\citep{rcv}.

The \pkg{STV} package implements the STV method \citep{STV}. 
The results are generally very similar to those from the \code{stv} 
function in the \pkg{vote} package. However, there are some minor 
differences that can lead to different results, particularly in elections 
with small electorates. Notably, in the \pkg{STV} package all quotas, 
vote counts and transfers are rounded to integers, which can lead to 
different results when the electorate is small. 
Also, in the \pkg{STV} package all tie-breaking is done at random, 
in contrast with the \pkg{vote} package, which uses forwards and backwards tie-breaking. 
Unlike the \pkg{vote} package, none of these other packages implements the STV method with equal ranking, or allows for reserved positions for marked groups.

The \pkg{HighestMedianRules} implements voting rules electing the candidate
with the highest median grade \citep{HighestMedianRules,Fabre2020}. 
The \pkg{electoral} and \pkg{esaps} packages compute various measures of electoral systems; in spite of their names, they do not implement electoral systems or voting rules \citep{electoral,esaps}.

\paragraph{Acknowledgements:} The research of Raftery and 
\v{S}ev\v{c}{\'{\i}}kov{\'a} was supported by NIH grant R01 HD070936.
The authors are grateful to Salvatore Barbaro and Brendan Murphy for
helpful discussions.


\bibliography{voting}



\end{document}